\documentclass[article,aps,twocolumn,showpacs,amsmath,amssymb,superscriptaddress,nofootinbib,floatfix]{revtex4-2}

\usepackage{graphicx}
\usepackage{dcolumn}
\usepackage{bm}
\usepackage[colorlinks=true,linkcolor=blue,citecolor=blue,urlcolor=blue]{hyperref}

\usepackage{array,booktabs,tabularx}
\usepackage[caption=false]{subfig}
\usepackage[USenglish]{babel}
\usepackage{braket}
\usepackage{xcolor}
\usepackage{amsfonts}
\usepackage{amssymb}
\usepackage[normalem]{ulem}
\usepackage{amsmath}

\begin{document}

 \title{First-order phase transition driven by competing charge-order fluctuations in 1\textit{T}'-TaTe\texorpdfstring{$_2$}{2}}

\author{S. K. Mahatha}
\thanks{These authors contributed equally to this work.}
\affiliation{UGC-DAE Consortium for Scientific Research, University Campus, Khandwa Road, Indore-452001, India}
\affiliation{Ruprecht Haensel Laboratory, Deutsches Elektronen-Synchrotron DESY, 22607 Hamburg, Germany}

\author{A. Kar}
\thanks{These authors contributed equally to this work.}
\affiliation{Donostia International Physics Center (DIPC), Paseo Manuel de Lardizábal, E-20018, San Sebastián, Spain}

\author{J. Corral-Sertal}
\thanks{These authors contributed equally to this work.}
\affiliation{Departamento de Física Aplicada, Universidade de Santiago de Compostela, E-15782 Campus Sur s/n, Santiago de Compostela, Spain}
\affiliation{CiQUS, Centro Singular de Investigacion en Quimica Biolóxica e Materiais Moleculares, Departamento de Quimica-Fisica, Universidade de Santiago de Compostela, Santiago de Compostela, E-15782, Spain}

\author{Josu Diego}
\thanks{These authors contributed equally to this work.}
\affiliation{Department of Physics, University of Trento, Via Sommarive 14, 38123, Povo, Italy}

\author{A. Korshunov}
\affiliation{Donostia International Physics Center (DIPC), Paseo Manuel de Lardizábal. 20018, San Sebastián, Spain}

\author{C.-Y. Lim}
\affiliation{Donostia International Physics Center (DIPC), Paseo Manuel de Lardizábal. 20018, San Sebastián, Spain}

\author{F. K. Diekmann}
\affiliation{Institut f\"{u}r Experimentelle und Angewandte Physik and Ruprecht Haensel Laboratory, Christian-Albrechts-Universit\"{a}t zu Kiel, 24098 Kiel, Germany}

\author{D. Subires}
\affiliation{Donostia International Physics Center (DIPC), Paseo Manuel de Lardizábal, E-20018, San Sebastián, Spain}
\affiliation{University of the Basque Country (UPV/EHU), Basque Country, Bilbao, 48080 Spain}

\author{J. Phillips}
\affiliation{Departamento de Física Aplicada, Universidade de Santiago de Compostela, E-15782 Campus Sur s/n, Santiago de Compostela, Spain}
\affiliation{Instituto de Materiais iMATUS, Universidade de Santiago de Compostela, E-15782 Campus Sur s/n, Santiago de Compostela, Spain} 

\author{T. Kim}
\affiliation{Diamond Light Source Ltd, Harwell Science and Innovation Campus, Didcot, OX11 0DE, United Kingdom}

\author{D. Ishikawa}
\affiliation{Materials Dynamics Laboratory, RIKEN SPring-8 Center, 679-5148 Japan}
\affiliation{Precision Spectroscopy Division, SPring-8 JASRI, 679-5198, Japan}

\author{G. Marini}
\affiliation{Department of Physics, University of Trento, Via Sommarive 14, 38123, Povo, Italy}

\author{I. Vobornik}
\affiliation{CNR-Istituto Officina dei Materiali (CNR-IOM), Strada Statale 14, 34149 Trieste, Italy}

\author{Ion Errea}
\affiliation{Donostia International Physics Center (DIPC), Paseo Manuel de Lardizábal, E-20018, San Sebastián, Spain}
\affiliation{Centro de Física de Materiales (CFM-MPC), CSIC-UPV/EHU, E-20018, San Sebastián, Spain}
\affiliation{Fisika Aplikatua Saila, Gipuzkoako Ingeniaritza Eskola, University of the Basque Country (UPV/EHU), E-20018, San Sebastián, Spain}

\author{S. Rohlf}
\affiliation{Institut f\"{u}r Experimentelle und Angewandte Physik and Ruprecht Haensel Laboratory, Christian-Albrechts-Universit\"{a}t zu Kiel, 24098 Kiel, Germany}

\author{M. Kall\"{a}ne}
\affiliation{Institut f\"{u}r Experimentelle und Angewandte Physik and Ruprecht Haensel Laboratory, Christian-Albrechts-Universit\"{a}t zu Kiel, 24098 Kiel, Germany}

\author{V. Bellini}
\affiliation{Istituto di Nanoscienze, Consiglio Nazionale delle Ricerche, 41125 Modena, Italy}

\author{A.Q.R. Baron}
\affiliation{ Materials Dynamics Laboratory, RIKEN SPring-8 Center, 679-5148 Japan}
\affiliation{Precision Spectroscopy Division, SPring-8 JASRI, 679-5198, Japan}

\author{Adolfo O. Fumega}
\affiliation{Department of Applied Physics, Aalto University, 02150 Espoo, Finland}

\author{A. Bosak}
\affiliation{European Synchrotron Radiation Facility (ESRF), BP 220, F-38043 Grenoble Cedex, France}

\author{V. Pardo}
\email{victor.pardo@usc.es}
\affiliation{Departamento de Física Aplicada, Universidade de Santiago de Compostela, E-15782 Campus Sur s/n, Santiago de Compostela, Spain}
\affiliation{Instituto de Materiais iMATUS, Universidade de Santiago de Compostela, E-15782 Campus Sur s/n, Santiago de Compostela, Spain} 

\author{K. Rossnagel}
\email{rossnagel@physik.uni-kiel.de}
\affiliation{Ruprecht Haensel Laboratory, Deutsches Elektronen-Synchrotron DESY, 22607 Hamburg, Germany}
\affiliation{Institut f\"{u}r Experimentelle und Angewandte Physik and Ruprecht Haensel Laboratory, Christian-Albrechts-Universit\"{a}t zu Kiel, 24098 Kiel, Germany}

\author{S. Blanco-Canosa}
\email{sblanco@dipc.org}
\affiliation{Donostia International Physics Center (DIPC), Paseo Manuel de Lardizábal, E-20018, San Sebastián, Spain}
\affiliation{IKERBASQUE, Basque Foundation for Science, 48013 Bilbao, Spain}

\date{September 2025}

\begin{abstract}
First-order phase transitions, characterized by a discontinuous change in the order parameter, are intriguing phenomena in condensed matter physics. However, the underlying, material-specific, microscopic mechanisms often remain unclear. Here, we unveil a high-temperature incommensurate charge-order precursor with the wave vector $\mathbf{q}^* = (0, \frac{1}{4}+\delta, \frac{1}{2})$ in the 1\textit{T}' phase of TaTe$_2$, which competes with fluctuating high-temperature Ta trimer bonding states at $\mathbf{q}_\mathrm{CO} =(0, \frac{1}{3}, 0)$. The precursor state follows the temperature dependence of the hidden incommensurability of the \textit{quasi}-1D nested Fermi surface. In contrast, the low-temperature commensurate charge order at $\mathbf{q}_\mathrm{CO}$, characterized by a charge disproportionation of the inequivalent Ta sites, appears to be driven by local chemical bonding. Dynamical lattice calculations identify an imaginary optical mode at $\mathbf{q}^*$, involving an in-plane vibration of the Ta atoms forming a chain-like structure that renormalizes below $T_\mathrm{CO}$. Our experimental and theoretical observations suggest that the controversial first-order phase transition, as captured by phenomenological Ginzburg-Landau theory, results from the competition between two order parameters: one involving Fermi surface nesting and the other involving local chemical bonding. 
\end{abstract}

\maketitle
\section{Introduction}
In a one-dimensional (1D) metal at half-filling, Peierls' theorem \cite{Peierls} states that translating parallel segments of the Fermi surface by a nesting vector of $q = 2k_\mathrm{F}$ drives the formation of a charge density wave (CDW), accompanied by a periodic lattice distortion (PLD), at a critical temperature $T_\mathrm{CDW}$, which breaks the translational invariance of the system. This second-order phase transition is signaled by divergences in the electronic susceptibility and correlation length at $q=q_\mathrm{CDW}$ and $T=T_\mathrm{CDW}$, respectively. However, first-order phase transitions involving the release or absorption of latent heat and a discontinuity in the first derivative of the free energy \cite{Stanley}, are also common in solids undergoing CDW formation. Prominent examples include manganites \cite{Mira_1999}, Fe$_3$O$_4$ (Verwey transition) \cite{Walz_2002}, ferroelectrics \cite{Shirane_1954,Sicron_1994}, transition metal dichalcogenides (TMDs) \cite{Wilson:1969ab,Wilson:1975aa}, and, more recently, kagome metals \cite{Ortiz_2019,Teng_2022,Arachchige_2022}. 

\begin{figure*}
    \centering
    \includegraphics[width=1.0\linewidth]{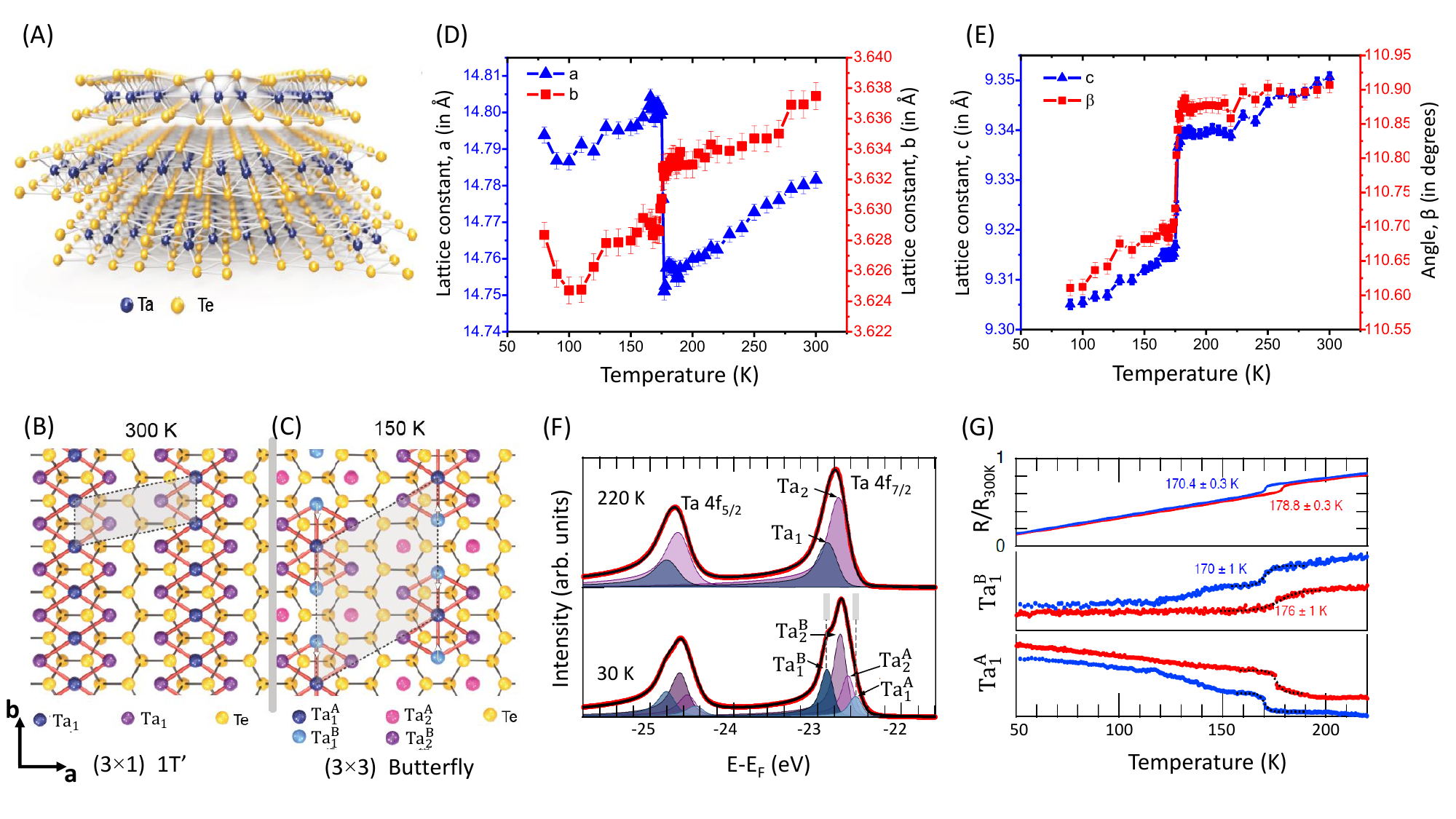}
    \caption{\textbf{Structure, lattice parameters, and charge disproportionation of 1\textit{T}'-TaTe$_2$}. \textbf{(A)} Lattice structure of 1\textit{T}'-TaTe$_2$. \textbf{(B-C)} High-temperature (HT) and low-temperature (LT) structures, which highlight the Ta atoms forming $(3\times1)$ zigzag chains and $(3\times3)$ butterfly-like structures, respectively. \textbf{(D-E)} Temperature dependence of the lattice parameters \textit{a}, \textit{b}, \textit{c}, and the monoclinic bond angle $\beta$, showing a jump at the transition temperature of $\sim$180\,K. \textbf{(F)} Core-level photoemission spectra of the spin-orbit split Ta 4\textit{f} emissions in the HT phase (220\,K) and LT phase (30\,K). In the HT phase, the spectral contribution results from Ta$_1$ and Ta$_2$ sites that split up into A and B sites at low temperatures. Vertical gray bars indicate the energy window used to track the temperature-dependent Ta$_{1}^\mathrm{A}$ and Ta$_{1}^\mathrm{B}$ photoemission intensity. \textbf{(G)} Comparison of the temperature-dependent normalized resistance and the Ta$_{1}^\mathrm{A}$ and Ta$_{1}^\mathrm{B}$ 4\textit{f} photoemission intensities.}
    \label{Fig1}
\end{figure*}

TMDs are a paradigmatic example of solids that exhibit first- and second-order CDW transitions \cite{Rossnagel_2011}. TMDs also host nematic phases \cite{Schrade_2024} and superconductivity \cite{Xi_2016}, making them an attractive platform to study the interplay between broken symmetry phases. Continuous CDW phase transformations, anticipated by high-temperature charge fluctuations, have been observed in TMDs, featuring the collapse of soft phonon modes \cite{Diego2021,Web11Nb,Web11Ti}, Fermi surface nesting \cite{Borisenko_2009}, and exciton condensation \cite{Cercellier_2007}. However, the microscopic origin of discontinuous (first-order) CDW transitions is far from being understood. These transitions typically imply strong-coupling theories, order-disorder scenarios, or competing phases \cite{Cowley_1979,Bruce_1979,Bruce_1979a}. One particular case is found in quasi-1D tellurium-based TMDs (MTe$_2$, M= V, Nb, Ta, Ir), which exhibit a first-order CDW transition \cite{Saleh:2020aa,Cao:2013aa} that presumably involves strong coupling of electrons to the cooperative distortions of the MTe$_6$ octahedra. Compared with their sulfur- and selenium-based counterparts, MTe$_2$ compounds with strong spin-orbit (SO) coupling undergo larger lattice distortions \cite{Katayama_2023} and charge transfer \cite{Whangbo:1992aa}, arising from the covalent Te--Te bonds between layers \cite{Canadell_1992}. These factors complicate the ground state and the understanding of the microscopic origin of CDW formation. Moreover, the non-integer number of electrons per site favors the formation of molecular orbitals (referred to as dimers, trimers, or heptamers) to minimize Coulomb repulsion at low temperatures \cite{Baggari:2020aa}. 

\begin{figure*}
    \centering
    \includegraphics[width=1.0\linewidth]{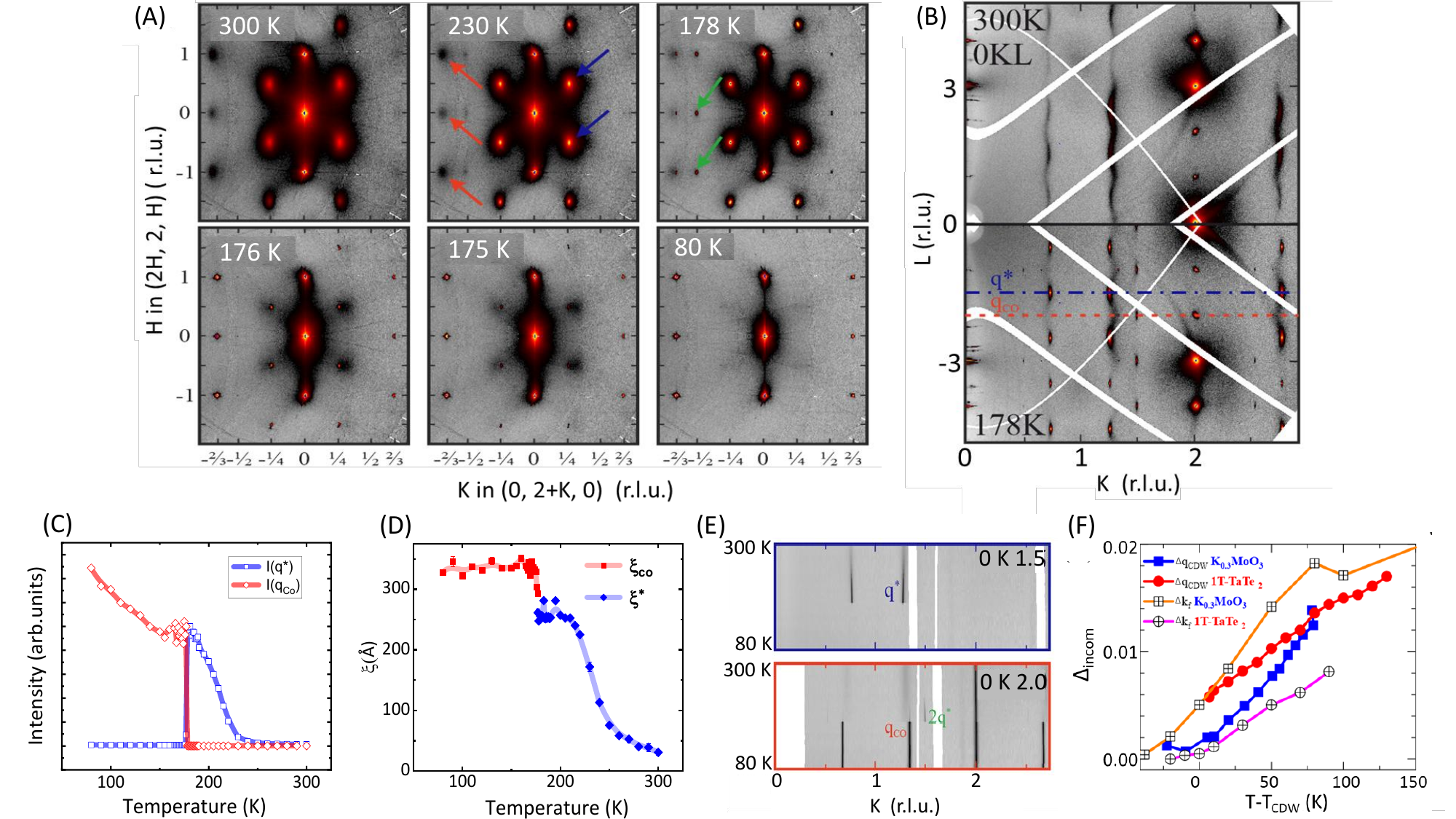}
    \caption{\textbf{x-ray diffuse scattering of 1\textit{T}'-TaTe$_2$}. \textbf{(A)} Temperature dependence of the $(2H, K, H)$ diffuse maps, highlighting the diffuse signals at $\mathbf{q}_\mathrm{CO}=(0, \frac{1}{3}, 0)$ (red arrows in the 230\,K panel) and $\mathbf{q}^*=(0, \frac{1}{4}+\delta, \frac{1}{2})$ (blue arrows) and the second harmonic of $\mathbf{q}^*$ (green arrows in the 178\,K panel). \textbf{(B)} Temperature dependence of the diffuse scattering intensity in the $(0, K, L)$ plane. Dashed red and blue lines show the $K$ dependence of $\mathbf{q}_\mathrm{CO}$ and $\mathbf{q}^*$, respectively. \textbf{(C)} Temperature dependence of the diffuse scattering intensity and \textbf{(D)} correlation lengths ($\xi$) of $\mathbf{q}_\mathrm{CO}$ and $\mathbf{q}^*$. \textbf{(E)} Waterfall plot displaying the temperature dependence of the charge-order precursors in \textbf{(A)} and \textbf{(B)}. \textbf{(F)} Temperature dependence of the incommensurability parameter $\Delta_\mathrm{incom} =q^*(T)-q^*(T_\mathrm{CO})$, as obtained from diffuse scattering, compared to the temperature dependence of the Fermi wavevector $k_{F_2}$. Also shown are corresponding results for quasi-1D K$_{0.3}$MoO$_3$.}
    \label{Fig2}
\end{figure*}

The 1\textit{T}' phase of TaTe$_2$ has recently garnered interest due to its topological properties \cite{Chen:2017ab}, large magnetoresistance \cite{Ali_2014}, and the $3\times1$ double zigzag chain-like structure (`ribbon chain') \cite{Brown:1966aa}, which is formed by a superposition of local fluctuating trimers \cite{Chen:2018aa}. Due to the lower chalcogen electronegativity, each Ta atom retains only $\frac{1}{3}$ of the electrons, resulting in a non-integer electron count per Ta site \cite{Vernes:1998aa,Canadell_1992}. This favors a tendency toward a $d^{1\pm\xi}$-electron count and stronger interlayer interactions that distort the ribbon chain. Below $T_\mathrm{CO}<180$\,K, a first-order charge order transition with a large temperature hysteresis occurs \cite{Liu:2015bb,Sorgel:2006aa,Chen:2017ab}. The PLD consists of a $(3\times3)$ \textit{butterfly}-like superstructure of metal atoms with short and long bonds. It is driven by the formation of Ta--Ta trimers due to significant local atomic displacement and orbital overlap during trimerization \cite{Gao:2018aa,Katayama_2023}. Contrary to the expected increase resulting from the opening of a CDW gap, a pronounced drop in resistivity occurs at $T_\mathrm{CO}$ \cite{Chen:2017ab}. Angle-resolved photoemission spectroscopy (ARPES) measurements suggest that the CDW transition is driven by quasi-1D nesting of the Fermi surface \cite{Lin:2022aa}, but the nesting vector does not match the wave vector of the charge modulation. In addition, scanning tunneling microscopy and spectroscopy revealed complex lattice distortions and intertwining between the $(3\times1)$ stripe and $(3\times3)$ charge modulations \cite{Feng:2016ab} that could mask a competing scenario favoring the observed first-order character of the transition. Consequently, the impact of the formation of these molecular orbitals on the lattice dynamics, the origin and mechanism of the discontinuous phase transition, and the anomalous temperature-dependent resistivity remain unclear. 

Here, we present a comprehensive experimental and theoretical study of the distorted 1\textit{T}' phase of TaTe$_2$ to understand its discontinuous first-order phase transition at the microscopic level. Our results show that the low-temperature $(3\times3)$ PLD with $\mathbf{q}_\mathrm{CO} =(0, \frac{1}{3}, 0)$, preceded by commensurate diffuse trimer fluctuations, results from chemical bonding instability that drives charge disproportionation at low temperatures. Moreover, we identify a temperature-dependent incommensurate charge-order precursor state with wave vector $\mathbf{q}^*=(0, \frac{1}{4}+\delta, \frac{1}{2})$ that competes with $\mathbf{q}_\mathrm{CO}$ at high temperatures. Remarkably, the incommensurability parameter $\delta$ maps the temperature dependence of the nested Fermi surface, similar to the nesting mechanism developed within the Peierls scenario. Inelastic x-ray scattering (IXS) spectra reveal a damped phonon at $\mathbf{q}^*$ and $\mathbf{q}_\mathrm{CO}$ that does not collapse at $T_\mathrm{CO}$. A phenomenological Ginzburg-Landau theory captures the first-order character of the phase transition, which is the result of the competition between the two order parameters. Thus, the high-temperature charge fluctuations are characterized by a collective interplay between dynamic molecular bonds $(3\times3)$ and nesting-driven incommensurate $(3\times4)$ orders. Our work contributes to the microscopic understanding of the discontinuous phase transition in 1\textit{T}'-TaTe$_2$, with implications for \textit{quasi}-1D materials, TMDs, and kagome metals.

\section{Results}
\subsection{x-ray diffraction and x-ray photoemission spectroscopy}

At high temperatures (HT), 1\textit{T}'-TaTe$_2$ is characterized by a distorted 2D structural motif resembling hexagonal 2D materials [Fig.~\ref{Fig1}(A)], although the unit cell is indexed within the monoclinic \textit{C2/m} space group \cite{Gao:2018aa,Vernes:1998aa}. The \textit{x}-\textit{y} projection reveals sheets of corner- and edge-sharing TaTe$_6$ octahedra that extend along the 1D \textit{b}-direction [Fig.~\ref{Fig1}(B)]. These sheets adopt a $(3\times1)$ double zigzag chain-like structure with two non-equivalent Ta atoms. Ta$_1$ atoms form the middle string, while Ta$_2$ atoms surround the outer slabs of the zigzag chain. Below the structural transition at $T_\mathrm{CO} = 180$\,K, clustering of Ta atoms lowers the symmetry by forming a \textit{butterfly}-like structure [Fig.~\ref{Fig1}(C)], which results from splitting the Ta$_1$ atoms into Ta$_{1}^\mathrm{A}$ and Ta$_{1}^\mathrm{B}$ sites. Consequently, the Ta$_{2}^\mathrm{B}$ atoms move toward the Ta$_{1}^\mathrm{A}$ atoms along the \textit{b}-axis \cite{Petkov:2020aa}. This creates a sequence of two short Ta$_{1}^\mathrm{A}$--Ta$_{1}^\mathrm{B}$ trimer molecular bonds \cite{Wang_2020} and one longer Ta$_{1}^\mathrm{B}$--Ta$_{1}^\mathrm{B}$ bond in the middle slab. 

Figures~\ref{Fig1}(D--E) show the temperature dependence of the lattice parameters and the monoclinic angle ($\beta$), as obtained from the refinement of x-ray diffraction data. Upon cooling, the lattice parameters and the monoclinic angle vary smoothly, followed by a discontinuous variation in unit cell volume at $T_\mathrm{CO}$ [see Supplementary Fig.~2(B)], consistent with the first-order character of the phase transition. However, the \textit{a}-lattice parameter, which accounts for the temperature dependence of the interchain distance, sharply increases at $T_\mathrm{CO}$, in agreement with the anomalous temperature dependence of the Ta--Ta distance at low temperatures \cite{Katayama_2023}.   

To learn more about the low-temperature (LT) phase of 1\textit{T}'-TaTe$_2$, we tracked the temperature dependence of the Ta 4\textit{f} emission using core-level photoemission spectroscopy (XPS), which is sensitive to the local charge density and atomic coordination. In the $(3\times1)$ HT phase ($T = 220 $\,K), Ta atoms occupy two different Wyckoff positions with a ratio of  Ta$_1$:Ta$_2$=1:2 per unit cell [Fig.~\ref{Fig1}(F)]. However, upon cooling below $T_\mathrm{CO}$, a shoulder appears at a higher binding energy due to the $(3\times3)$ LT lattice reconstruction containing four inequivalent Ta sites (atomic ratio Ta$_{1}^\mathrm{A}$: Ta$_{1}^\mathrm{B}$: Ta$_{2}^\mathrm{A}$: Ta$_{2}^\mathrm{B}$= 1:2:2:4). Accordingly, the Ta 4\textit{f} XPS peaks were fitted with two and four Lorentzians above and below $T_\mathrm{CO}$, respectively, yielding the values listed in Table~\ref{Table}. These values can be compared with \textit{ab initio} calculations by tracking the shift in Ta core levels caused by the change in the Ta environment resulting from charge disproportionation. Using the low-energy peak as a reference (zero), we observe that the calculated shift in the inequivalent Ta 1\textit{s} levels aligns with the XPS results (see Table~\ref{Table}), which is consistent with previous reports on charge-ordered nickelates \cite{quan2012formal}.

\begin{figure*}
    \centering
    \includegraphics[width=1.0\linewidth]{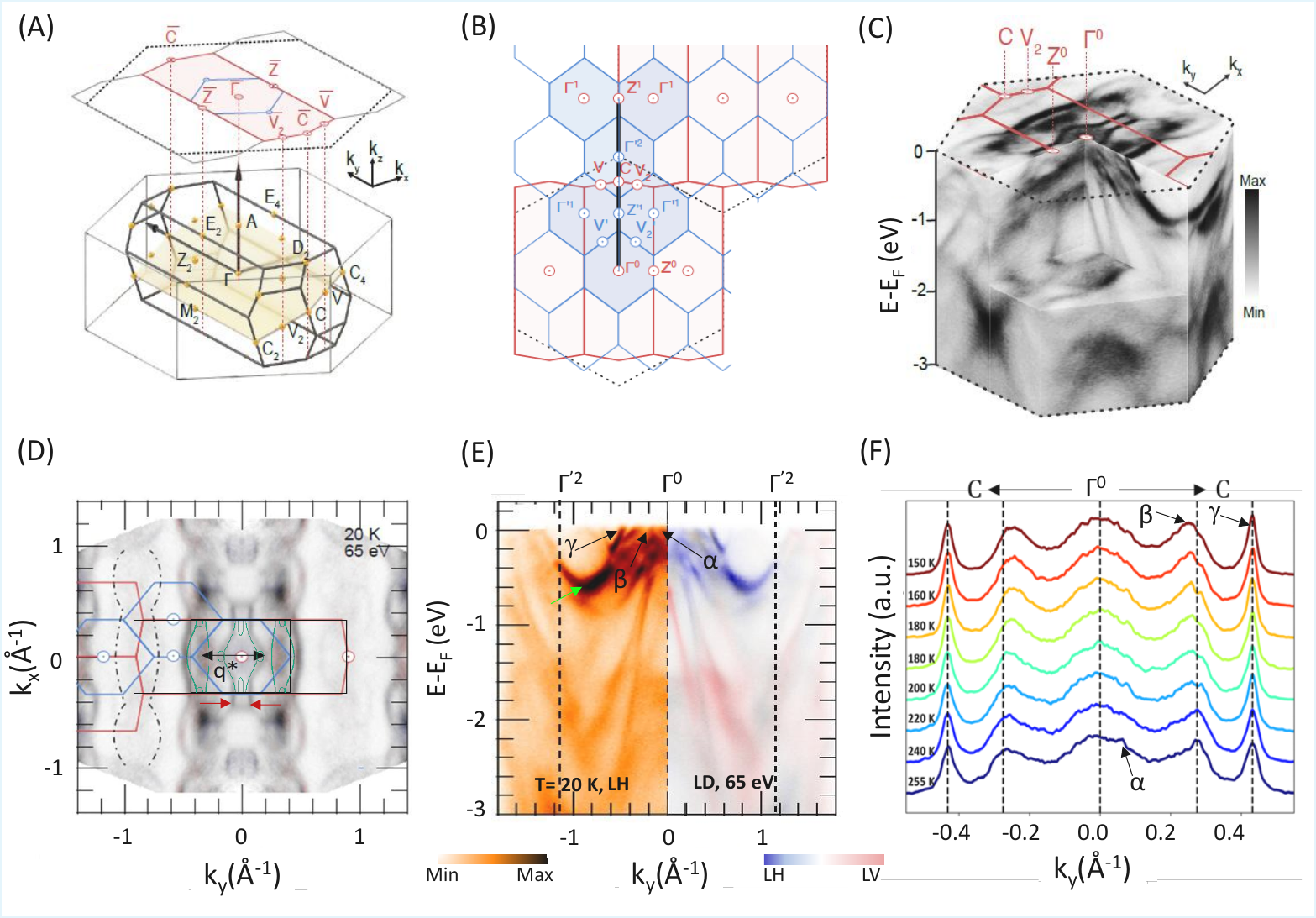}
    \caption{\textbf{Angle-resolved photoemission spectroscopy of 1\textit{T}'-TaTe$_2$}. (\textbf{A-B}) Bulk and surface-projected Brillouin zones, as well as the high-symmetry directions of the undistorted $(1\times1)$ phase (black dashed hexagons), the HT $(3\times1)$ phase (red elongated hexagons), and the LT $(3\times3)$ phase (blue hexagons). The relevant $\Gamma^0$--$Z^1$ directions are indicated by the blue mini-Brillouin zones. (\textbf{C}) Experimental 3D Fermi surface at 20\,K obtained using an incident photon energy of 65\,eV. (\textbf{D}) $k_x$--$k_y$ projection of the experimental Fermi surface taken at 20\,K ($h\nu=65$\,eV), overlaid with the DFT-calculated Fermi surface. The red circles indicate the high-symmetry points of the HT $(3\times1)$ surface Brillouin zone, and the blue hexagon represents the LT $(3\times3)$ Brillouin zone. (\textbf{E}) Valence band (left panel)and linear dichroism (right panel) ($\mathrm{LD} = \mathrm{LH}-\mathrm{LV}$) ARPES spectra along the $\Gamma$--$Z$ symmetry direction at 20\,K using 65\,eV photons. Blue and red represent the spectral intensity obtained for linear horizontal (LH) and linear vertical (LV) polarized light, respectively. (\textbf{F}) Temperature-dependent evolution of the band spectrum crossing the Fermi level around the zone center along the $\Gamma^0$--$C$ symmetry direction.}
    \label{Fig3}
\end{figure*}

\begin{table}[htbp]
\global\long\def\arraystretch{1.12}
\caption{Extracted peak positions of inequivalent Ta atoms and asymmetries ($\alpha$) of the photoemission lineshapes for the HT and LT phases of 1\textit{T}'-TaTe$_2$. The uncertainty in the peak position is $\Delta E\sim5$\,meV. DFT energy differences were calculated for the Ta 1\textit{s} core levels and referenced to the Ta$_2^\mathrm{A}$ 1\textit{s} energy.}

\setlength{\tabcolsep}{0.6mm}{
\begin{tabular}{c|c|c|c|c|c|c|c|c|}
\hline
\hline

  &     220 K     &     90 K    &    DFT-HT  & DFT-LT \\
  \hline
  
     E$_\mathrm{Ta1A}$ (eV)     &    -22.85(8)  &  -22.87(4)  & 0.09   &  0.24  \\
     E$_\mathrm{Ta1B}$ (eV)     &      &  -22.73(3)  &    & 0.15  \\
     E$_\mathrm{Ta2A}$ (eV)  &    -22.72(3)  &  -22.62(4)  & 0.00  & 0.00 \\
     E$_\mathrm{Ta2B}$ (eV)     &      &  -22.71(7)  &    &  0.15\\
     $\alpha$ (eV)$^{-1}$  &      0.190  &  0.165  &  & \\
\hline
\hline
\end{tabular}}
\label{Table}
\end{table}

Additionally, the temperature dependence of the integrated intensity of the Ta$_{1}^\mathrm{A}$ and Ta$_{1}^\mathrm{B}$ 4\textit{f} emissions shows a sharp drop around $170\pm1$\,K with $\sim$6\,K hysteresis. This temperature behavior follows that of the normalized resistance [Fig.~\ref{Fig1}(G)] and LEED superstructure reflections (see Supplementary Sec.~II). The observed charge disproportionation of the Ta$_{1}^\mathrm{A}$ and Ta$_{1}^\mathrm{B}$ atoms below $T_\mathrm{CO}$ is consistent with a chemical bonding instability that accompanies the first-order phase transition, as reported for IrTe$_2$ \cite{Ko_2015}.

\subsection{x-ray diffuse scattering}

Figure \ref{Fig2} provides an overview of the temperature dependence of diffuse scattering (DS) data collected for 1\textit{T}'-TaTe$_2$. Due to its low dimensionality and weak interlayer interactions, fluctuations of the trimer molecular bonds are anticipated in the temperature interval between the mean field (MF) transition and the charge order (CO) transition, $T_\mathrm{CO} < T < T_\mathrm{MF}$. At room temperature (RT), diffuse intensity clouds at the wave vector $\mathbf{q}_\mathrm{CO} =(0, \frac{1}{3}, 0)$, indicated by the red arrows in Fig.~\ref{Fig2}(A), appear around the $(0, \overline{2}, 0)$ Bragg position. The DS unambiguously identifies the fluctuating charge-order precursor of the trimer molecular bonds, in agreement with the refinement of the x-ray diffraction data and the literature \cite{Gao:2018aa,Katayama_2023}. The diffuse intensity clouds are short-range correlated in the $(2H, K, H)$ plane, with a correlation length $\xi\approx 10$\,\r{A}, and evolve toward a long-range charge order with $\xi_\mathrm{CO} \approx 350$\,\r{A} at 180\,K, as a consequence of the Ta--Ta charge disproportionation that triples the \textit{b}-axis parameter below $T_\mathrm{CO}$ [Fig.~\ref{Fig2}(D)]. Interestingly, the intensity of the charge-order precursor decreases upon cooling and increases sharply again upon approaching $T_\mathrm{CO}$ [Fig.~\ref{Fig2}(C)], contradicting the expected behavior of thermal excitation of a soft mode as a phase transition is approached. Below $T_\mathrm{CO}$, the integrated intensity grows smoothly, while the correlation length remains constant down to 100\,K. 

Intriguingly, the DS associated with the $(3\times3)$ PLD coexists with a strong incommensurate diffuse signal at $\mathbf{q}^*=(0, \frac{1}{4}+\delta, \frac{1}{2})$, as indicated by the blue and green arrows in Fig.~\ref{Fig2}(A). At both $\mathbf{q}_\mathrm{CO}$ and $\mathbf{q}^*$, the signals develop a rod-like 2D structure perpendicular to the TaTe$_2$ layers, consistent with decoupled out-of-plane CDW fluctuations [Fig.~\ref{Fig2}(B)]. Upon cooling from RT to $T_\mathrm{CO}$, the 2D charge fluctuations begin to condense at $L=\frac{1}{2}$, exhibiting a 2D-to-\textit{quasi}-3D crossover. Conversely, the DS intensity at $\mathbf{q}^*$ increases with decreasing temperature, but is less correlated than the $(3\times3)$ PLD ($\xi^* \approx 250$\,\r{A} at 180\,K) and is sharply suppressed at $T_\mathrm{CO}$ [Fig.~\ref{Fig2}(D)]. The opposite trend in the temperature dependence of the integrated DS intensities at both $\mathbf{q}_\mathrm{CO}$ and $\mathbf{q}^*$ hints at their mutual competition, similar to recent reports on kagome metals \cite{subires2025frustrated,Cao_2023} and high-$T_c$ cuprates \cite{Kim_2018}. The incommensurability parameter, $\Delta_\mathrm{incom} = q^*(T) - q^*(T_\mathrm{CO})$, revealed in the waterfall plot in Fig.~\ref{Fig2}(E) and plotted explicitly in Fig.~\ref{Fig2}(F), decreases with decreasing temperature and retains a residual incommensurability of $\Delta_\mathrm{incom}=0.06$\,\r{A}$^{-1}$ below $T_\mathrm{CO}$, similar to what was observed for the \textit{quasi}-1D blue bronze K$_{0.3}$MoO$_3$ \cite{Fleming_1985,Fedorov_2000}.

\subsection{Angle-resolved photoemission spectroscopy}

The idea that 1\textit{T}'-TaTe$_2$ shares the same hidden charge-order precursor as K$_{0.3}$MoO$_3$ can be tested further experimentally by ARPES. Figure \ref{Fig3}(C) summarizes the measured LT Fermi surface (FS) and valence-band dispersion in a 3D plot. The main symmetry directions in the HT and LT phases are indicated in Figs.~\ref{Fig3}(A) and \ref{Fig3}(B). At LT ($T=20$\,K), the high-resolution FS shows a quasi-one-dimensional double \textit{zigzag} chain with opposite curvature along \textit{k$_x$} [Fig.~\ref{Fig3}(D)], which is in good agreement with the overlaid DFT-calculated FS. Examining the valence-band spectra in detail [left panel in Fig.~\ref{Fig3}(E)], we identify three bands dispersing toward $E_\mathrm{F}$, perpendicular to the chain direction, and crossing at the Fermi vectors $k_{\mathrm{F}_1}$ ($\alpha$), $k_{\mathrm{F}_2}$ ($\beta$) and $k_{\mathrm{F}_3}$ ($\gamma$). The hole pockets at the $\Gamma$ and $\Gamma'$ points are connected by a linear band running parallel to the $k_x$ direction [red arrows in Fig.~\ref{Fig3}(D)]. Furthermore, we unveil a $(3\times3)$ FS reconstruction along the $\Gamma$--$C$ direction that gives rise to weak replica bands in the region between two parallel chain-like FS sheets [black dashed lines in Fig.~\ref{Fig3}(D)]. 

As shown in the right panel of Fig.~\ref{Fig3}(E), the linear dichroism demonstrates that the bands close to $E_\mathrm{F}$ are primarily derived from the hybridization of the out-of-plane Ta $d_{xz}$ and $d_{yz}$ orbitals, with a minor contribution from the out-of-plane Te $p_z$ orbitals (Supplementary Fig.~5). The experimental data reveal a V-shaped, downward-dispersing band with a minimum at $–0.6$\,eV [highlighted by the green arrow in the left panel of Fig.~\ref{Fig3}(E)], exhibiting strong spectral weight from the Te in-plane $p_x$ and $p_y$ orbitals. These results align well with DFT calculations that reproduce the ARPES spectra (see Supplementary Sec.~IV and Supplementary Figs.~4 and 5) and are consistent with previous reports in the literature \cite{Mitsuishi_2024, Lin:2022aa}. Furthermore, comparing the HT ($T = 220$\,K) and LT valence-band spectra reveals a reconstruction of the band structure over a binding energy range of 1\,eV (Supplementary Fig.~4). At LT, the $\gamma$ band crossing the Fermi level in Fig.~\ref{Fig3}(E) opens a 100-meV gap at $E_\mathrm{B} \approx -0.25$\,eV resulting from the avoided crossing of the band folding (Supplementary Fig.~4). 

Importantly, we unveil several possible nesting vectors that connect the warped FS and the two linear flat bands (see Supplementary Fig.~7 for a complete identification of the nesting vectors). Of these, the wave vector $\mathbf{q}^*$ connecting the bands with Fermi vectors $k_{\mathrm{F}_1} \approx -0.14$\,\r{A}$^{-1}$ ($\alpha$ band) and $k_{\mathrm{F}_2} \approx 0.26$\,\r{A}$^{-1}$ ($\beta$ band) in Fig.~\ref{Fig3}(E), matches the length of the wavev ector of the incommensurate charge-order precursor ($\sim$$0.43$\,\r{A}$^{-1}$). Remarkably, in the temperature interval between $T_\mathrm{CO}$ and RT, the momentum distribution curves (MDCs) along the $\Gamma$--$C$ direction display a temperature dependence of the Fermi wave vector $k_{\mathrm{F}_2}$ ($\beta$ band) [Fig.~\ref{Fig3}(F)]. While $k_{\mathrm{F}_3}$ of the $\gamma$ band does not change as a function of temperature, $k_{\mathrm{F}_2}$ shifts from $0.255$\,\r{A}$^{-1}$ at 255\,K to $0.275$\,\r{A}$^{-1}$ at 150\,K. We cannot exclude the possibility of a slight temperature dependence of $k_{\mathrm{F}_1}$ beyond our momentum resolution. However, the temperature dependence of the incommensurability parameter of the $\beta$ band parallels the slope of the DS precursor at $(0, \frac{1}{4}+\delta, \frac{1}{2})$ [Fig.~\ref{Fig2}(E)], thereby demonstrating that $q^*(T)$ follows the hidden dependence of the nested Fermi wave vectors $k_{\mathrm{F}_1}$ and $k_{\mathrm{F}_2}$. Nevertheless, identifying a nesting vector in the experimental ARPES spectra does not produce a peak in the calculated nesting function, even though the value of the electronic susceptibility is anomalously large as a function of momentum (Supplementary Sec.~VI and Supplementary Fig.~8). Furthermore, we do not find any parallel segments in the FS that match the wave vector $\mathbf{q}_\mathrm{CO}$ of the LT $(3\times3)$ phase (Supplementary Fig.~8).

\begin{figure*}
    \centering
    \includegraphics[width=1.0\linewidth]{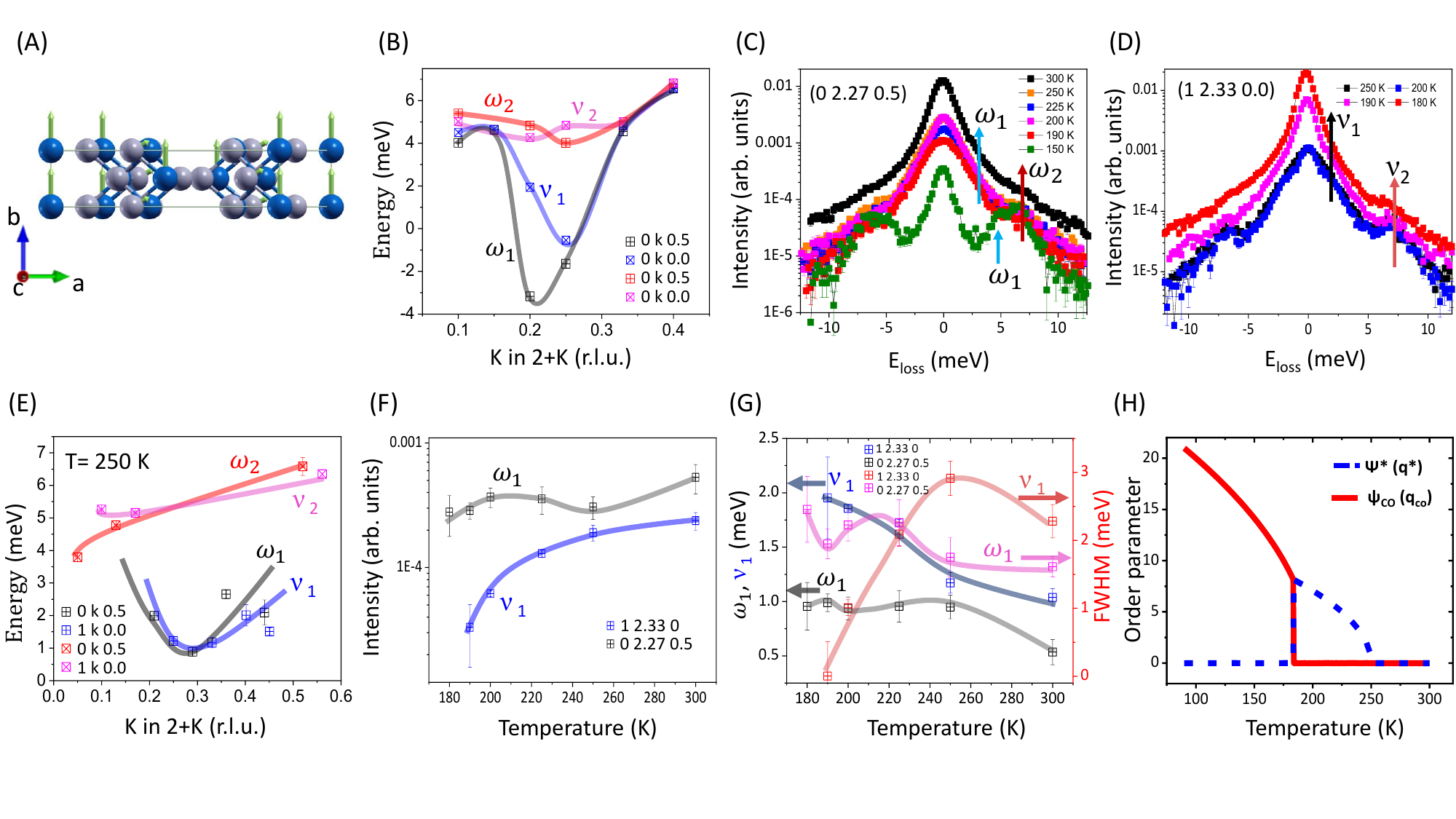}
    \caption{\textbf{Inelastic x-ray scattering of 1\textit{T}'-TaTe$_2$}. (\textbf{A}) Unit cell of 1\textit{T}'-TaTe$_2$ indicating Ta-atom vibrations at $\mathbf{q}^*$. (\textbf{B}) Theoretical harmonic phonon dispersion of the low-energy modes showing imaginary modes around $\mathbf{q}^*$ and $\mathbf{q}_\mathrm{CO}$. (\textbf{C}) Temperature dependence of the low-energy phonon modes $\omega_1$ and $\omega_2$ at $\mathbf{q}^*$. (\textbf{D}) Temperature dependence of $\nu_1$ and $\nu_2$ phonons at $\mathbf{q}_\mathrm{CO}$. (\textbf{E}) Wave-vector dependence of $\omega_1$ and $\nu_1$ at 250\,K. Corresponding temperature dependence of (\textbf{F}) the integrated phonon intensity and (\textbf{G}) the energy and full width at half maximum (FWHM). (\textbf{H}) Temperature dependence of the competing order parameters associated with $\mathbf{q}_\mathrm{CO}$ and $\mathbf{q}^*$ according to a minimal Ginzburg--Landau model.}
    \label{Fig4}
\end{figure*}

\subsection{Density functional theory and inelastic x-ray scattering}

Now that we have examined the crystal and electronic structure of 1\textit{T}'-TaTe$_2$ and identified the fluctuating precursors of the charge modulations, we will explore the lattice response to the competing orders. The zero-temperature harmonic DFT phonon band structure displays a negative optical branch ($\omega_1$) at the wave vector $\mathbf{q}^*$ matching the precursor observed in DS [Fig.~\ref{Fig4}(B) and Supplementary Fig.~10]. This indicates that the $(3\times1)$ structure is unstable toward a $(3\times4)$ reconstruction at LT. The imaginary mode at $\sim$$\mathbf{q}^*$ describes an in-plane vibration of Ta atoms along the chain direction [Fig.~\ref{Fig4}(A)]. Additionally, the mode $\nu_1$ at $(0, 0.25, 0)$ exhibits a shallow imaginary energy of approximately $–0.5$\,meV. While this suggests that the $(3\times4)$ PLD is not strictly stable at the harmonic level, the small magnitude of the instability implies that anharmonic lattice effects might be sufficient to stabilize the structure at LT. Conversely, the absence of negative energies at $\mathbf{q}_\mathrm{CO}$ = (0, $\frac{1}{3}$, 0) indicates that the $(3\times3)$ PLD is stable within the harmonic approximation. The phonon band structure of a hypothetical hexagonal 1\textit{T} structure exhibits a more pronounced instability at $(\frac{1}{3}, 0, 0)$ that naturally leads to the distorted 1\textit{T}' phase at HT (Supplementary Fig.~9). On the other hand, the significant renormalization of the optical phonon branch with electronic temperature (Supplementary Fig.~10) suggests that FS nesting and electron--phonon interaction (EPI) primarily drive the incommensurate charge-order precursor, despite the lack of a clear divergence in the real part of the electronic susceptibility (Supplementary Fig.~8).

From an experimental standpoint, the IXS spectra at $\mathbf{q}^*$ ($\mathbf{q}_\mathrm{CO}$) reveal an elastic central peak (CP) at an energy loss ($E_\mathrm{loss}$) of 0\,meV and two Stokes--anti-Stokes phonon branches labeled $\omega_1$ ($\nu_1$) and $\omega_2$ ($\nu_2$), respectively [Figs.~\ref{Fig4}(C--D)]. The IXS spectra were fitted to damped harmonic oscillations convoluted with the instrumental resolution ($\Delta E = 1.5$\,meV) and two phonon branches with energies of $\omega_1$ and $\nu_1$ $\approx1$\,meV and $\omega_2$ and $\nu_2$ $\approx7$\,meV (see Supplementary Sec.~VIII and Supplementary Figs.~11--15 for fitting details). First, we note that the elastic CP at $\mathbf{q}^*$ increases upon cooling and is suppressed at $T_\mathrm{CO}$, while at $\mathbf{q}_\mathrm{CO}$, it decreases progressively upon cooling and increases finally at $T_\mathrm{CO}$, in agreement with the DS results of Fig.~\ref{Fig2}. Moreover, the experimental phonon dispersion in the wave vector range of $0 < 2+K < 0.5$ r.l.u.\ in Fig.~\ref{Fig4}(E) contains a mixture of the spectral weights of the $\omega_1$ and $\omega_2$ modes. Within the range of $0 < 2+K < 0.2$ r.l.u., the low-energy spectra are dominated by the $\omega_2$ mode, while in the range of $0.2 < 2+K < 0.4$ r.l.u.\, the $\omega_1$ mode takes over the spectral dominance [Fig.~\ref{Fig4}(E) and Supplementary Fig.~16]. Furthermore, the intensity of the $\omega_1$ and $\nu_1$ modes decreases with temperature, indicating that the increase of the elastic CP mainly drives the temperature dependence of the DS [Fig.~\ref{Fig4}(F)]. This is further corroborated by the hardening of the $\nu_1$ mode at $\mathbf{q}_\mathrm{CO}$ upon cooling, which is consistent with the absence of an imaginary mode in the DFT calculations. 

However, the energies of the $\omega_1$ and $\nu_1$ modes at $\mathbf{q}^*$ and $\mathbf{q}_\mathrm{CO}$ are significantly lower (by about 1--2\,meV) than the $\sim$4--6\,meV predicted \textit{ab initio} in the noninteracting calculation ( Supplementary Fig.~10). The anomalously low energy of the $\omega_1$ mode is temperature independent below 250\,K and becomes damped over a broad range of wave vectors as it approaches $T_\mathrm{CO}$, presumably due to renormalization by EPI. Below $T_\mathrm{CO}$, the energy of the $\omega_1$ mode increases up to $\sim$5\,meV [cyan arrow in Fig.~\ref{Fig4}(C)], while its spectral weight is largely suppressed due to the modification of the dynamical structure factor in the LT phase. The significant enhancement of the elastic line at $\mathbf{q}_\mathrm{CO}$ precludes accurately determining the energy of $\nu_1$ below $T_\mathrm{CO}$. The temperature dependence of the $\nu_1$ mode demonstrates that the fluctuating molecular bonds in \textit{quasi}-1D TMDs do not fit within a soft mode scenario \cite{Web11Nb,Web11Ti,Diego2021}. Conversely, the $\omega_1$ spectra display continuous broadening upon cooling, which contradicts the scenario in which anharmonicity decreases due to phonon--phonon interactions, while the spectra of the $\nu_1$ mode narrow significantly and become resolution-limited at $T_\mathrm{CO}$. The strong phonon anharmonicity that we reveal here for the $\omega_1$ mode is induced by the coupling of electronic and vibrational degrees of freedom. However, neither the FS nesting or EPI nor bond instability of the charge-order precursors at $\mathbf{q}^*$ and $\mathbf{q}_\mathrm{CO}$ drive freezing of a soft mode in the low-symmetry phase.

\section{Discussion}

The large set of experimental data presented here, supported by DFT and phonon calculations, enables us to develop a detailed and comprehensive understanding of the phase transition in 1\textit{T}'-TaTe$_2$. We identified fluctuating trimer bonds at HT with a wave vector of $(0, \frac{1}{3}, 0)$ as the precursors of the charge order at LT. These fluctuations turn into the $(3\times3)$ PLD, which is characterized by a charge disproportionation of the Ta sites. Furthermore, the DS and ARPES experiments identify nesting-driven charge fluctuations with wave vector $\mathbf{q}^*$ that compete with chemical bond instability and eventually give rise to the first-order character of the phase transition. As the phase transition is approached, the DS data show that the crystal evolves toward a commensurate phase at $\mathbf{q}^*$, but, instead, the LT commenurate charge-disproportionated state forms. The competition between charge ordering at $\mathbf{q}_\mathrm{CO}$ and $\mathbf{q}^*$ is well captured by a phenomenological Ginzburg--Landau theory that incorporates the coupling of two order parameters in the free energy expansion:
\begin{equation}
    \begin{split}
    F = a_1(T)|\psi_\mathrm{CO}|^2 +  \frac{b_1}{2}|\psi_\mathrm{CO}|^4 + a_2(T)|\psi^{*}|^2 + \\ 
    \frac{b_2}{2}|\psi^{*}|^4 + \gamma |\psi_\mathrm{CO}|^2 |\psi^{*}|^2
    \end{split}
\label{GL}
\end{equation}
Here, $\psi_\mathrm{CO}$ and $\psi{^*}$ represent the amplitude of the order parameters associated with charge orders at $\mathbf{q}_\mathrm{CO}$ and $\mathbf{q}^*$, respectively, and $\gamma=1$ stands for a strong (competing) coupling term between the two order parameters, and thereby characterizes the phase transition (see Supplementary Sec.~X). Figure~\ref{Fig4}(H) displays the temperature dependence of the order parameters $\psi_\mathrm{CO}$ and $\psi{^*}$. We note that the instability at $\mathbf{q}^*$ is the primary order parameter, and its interaction lowers the transition temperature for $\mathbf{q}_\mathrm{CO}$, thus developing a first-order phase transition. This is in agreement with the experimental results. The competition between the two order parameters precludes softening of the low-energy mode at $\mathbf{q}^*$, provided this coupling is strong enough. However, the $\omega_1$ phonon becomes anharmonic due to EPI. These observations depict a charge-ordering mechanism that differs from simple FS nesting or EPI scenarios, with the anharmonic phonon occurring at $\mathbf{q}^*$, while the static charge order emerges at the distinct wave vector $\mathbf{q}_\mathrm{CO}$. First-principles calculations reveal that formation of charge order at $\mathbf{q}^*$ is energetically more favorable, as corroborated experimentally by the DS intensity at HT. However, DFT and XPS demonstrate that the large energy gain due to charge disproportionation, preempted by the fluctuating trimer molecules at HT, promotes charge order at $\mathbf{q}_\mathrm{CO}$ as the ground state.    

The enigmatic drop in resistivity at $T_\mathrm{CO}$ fits within this scenario. The linear temperature dependence of the resistivity at $T>T_\mathrm{CO}$ demonstrates an electron transport dominated by the scattering with the anharmonic $\omega_1$ mode. The strong renormalization of the $\omega_1$ mode below $T_\mathrm{CO}$, which increases its energy from $\sim$1\,meV to $\sim$5\,meV [Fig.~\ref{Fig4}(C)], removes scattering channels from the Fermi level. Therefore, a large resistivity drop occurs upon cooling below $T_\mathrm{CO}$.
Furthermore, a full gap opening associated with either $\mathbf{q}^*$ or $\mathbf{q}_\mathrm{CO}$ is unlikely to occur in 1\textit{T}'-TaTe$_2$, consistent with transport measurements.

To conclude, we will place our results in a broader context within the family of 1D metals and first-order CO phase transitions. A temperature-dependent nesting wave vector has also been reported for \textit{quasi}-1D K$_{0.3}$MoO$_3$ ($T_\mathrm{CDW}=180$\,K), which develops a LT CDW at the wave vector $\mathbf{q}_\mathrm{CDW} =(0, \frac{1}{4}+\delta, \frac{1}{2})$. The degree of incommensurability of this wave vector is similar to that of the temperature-dependent Fermi wave vectors, which lock to a commensurate value at 100\,K. This is analogous to the hidden dependence of the FS that we have revealed here, suggesting a common mechanism in the blue bronzes and TMDs. However, despite the length of the nesting wave vector in 1\textit{T}'-TaTe$_2$ being very close to the $\sim$$\frac{1}{4}$ value at 20\,K, it remains incommensurate, presumably caused by phonon dispersions or interchain interactions \cite{Noguera_1991}. 

Our results also share similarities with the charge order observed in rare earth nickelates \cite{Anissimova_2014}, manganites, and, more specifically, in PbZrO$_3$ and magnetite \cite{Wright_2001}. These materials exhibit order--disorder transformations without phonon softening despite the strong coupling of electronic, orbital, and vibrational degrees of freedom. The antiferroelectric transition in PbZrO$_3$ has been explained by the coupling of two order parameters corresponding to two different points in the BZ, featuring a soft mode that is prevented by their repulsive interaction \cite{Tagantsev_2013,Xu_2019}. Similarly, the celebrated first-order Verwey phase transition in Fe$_3$O$_4$, characterized by a LT charge modulation and anticipated by HT fluctuating trimer bond molecules, is driven by the coupling of two order parameters. One of these exhibits a strong anharmonicity in the low-energy phonon spectra at the transition temperature \cite{Piekarz_2006,Hoesch_2013}.
  
In conclusion, we have presented a comprehensive set of experimental and theoretical data supporting a framework of competing order parameters to explain the first-order phase transition and the controversial temperature-dependent behavior of the 1\textit{T}' phase of TaTe$_2$. Our results uncover a novel mechanism to explain phase transitions in quasi-2D TMDs, beyond the conventional second-order phase transition. They also reveal a competing scenario of order parameters, as observed in superconducting cuprates and the multiple $q_\mathrm{CDW}$ observed in kagome metals.

\section{Methods}
High-quality single crystals of 1\textit{T}'-TaTe$_2$ were grown by chemical vapor transport (temperature gradient of 905 to 800 ºC) using iodine as a transport agent for 6 weeks. The crystals had a typical size of a few mm in the form of elongated plates with the long dimension parallel to the short \textit{b}-axis. Electrical resistance measurements were carried out in the standard four-probe geometry using a Quantum Design physical property measurement system. Residual resistance ratios were in the range of R(300 K)/R(4 K) $\sim$ 20.4 $\pm$ 0.3, Supplementary Fig. 2(A). Temperature-dependent LEED (Supplementary Fig. 3) as well as x-ray photoelectron spectroscopy (XPS) and soft x-ray angle-resolved photoemission spectroscopy (ARPES) measurements were done at the ASPHERE III endstation at beamline P04 of PETRA III, DESY, Hamburg. LEED patterns were recorded at an electron kinetic energy of 80 eV. Photoemission spectra were acquired in a photon energy range of 260 eV-525 eV using a Scienta-Omicron DA 30 spectrometer (Supplementary Fig. 4). The energy and momentum resolutions were about 40 meV and 0.02 \r{A}$^{-1}$, respectively. The photon beam size on the sample was 10$\times$30 $\mu$m$^2$ with a total photon flux of almost 10$^{13}$ photons/second. An appropriate line-shape model for the Ta 4\textit{f} shallow core-levels was used, accounting for the phonon broadened core-hole lifetime and a slow electron tail due to a possible energy loss that is represented by the excitation model J8 from Hughes and Scarfe for a least-square fit.

ARPES measurements in the vacuum-ultraviolet photon energy range were performed at the APE-LE beamline of Elettra synchrotron in Trieste using linear horizontal and vertical polarized light. All photoemission measurements were acquired using a photon energy of 65 eV and a Scienta Omicron DA30-L analyzer with a total energy resolution of 40 meV. The photon beam size on the sample was 50$\times$150 $\mu$m$^2$. ARPES experiments at DIAMOND were carried out at the high resolution branch (HR-ARPES) \cite{Hoesch_RSI_2017} of the beamline I05 in the Diamond Light Source, equipped with an MBS A-1 analyzer. Samples were in-situ cleaved under vacuum better than 2$\times$10$^{-10}$ mbar. Temperature-dependent valence band measurements were performed using a photon energy of h$\nu$ = 65 eV. 

Single crystal diffuse scattering was performed at the ID28 beamline at ESRF with \textit{E}$_i$=17.8 keV and a Dectris PILATUS3 1M X area detector. We use the CrysAlis software package for the orientation matrix refinement and the ID28 software ProjectN for reconstruction of the reciprocal space maps, and plotted them in Albula. The components ($h$ $k$ $l$) of the scattering vector are expressed in reciprocal lattice units (r.l.u.), ($h$ $k$ $l$)= $h\ \mathbf{a}^*+k \mathbf{b}^*+l\ \mathbf{c}^*$, where $\mathbf{a}^*$, $\mathbf{b}^*$, and $\mathbf{c}^*$ are the reciprocal lattice vectors. The IXS experiments were conducted at the BL43LXU \cite{Baron2019,Baron2020} of SPring-8 using backscattering at the Si(11 11 11) reflection at 21.75 keV in May of 2023. The incident flux was about 2x10$^10$ photons/s in a bandwidth of $\sim$0.8 meV and spot size of $\sim$50x50 $\mu$m$^2$ full width at half maximum (FWHM). Most measurements used one analyzer of the available 4x7 array, with an over-all energy resolution 1.4 meV FWHM. Scans were typically done over a range of $\pm$ 12 meV, with one scan requiring about 40 minutes. The momentum transfer resolution was set by 30x30 mm$^2$ slits located 9 meters from the sample, corresponding to ($\Delta$H, $\Delta$K, $\Delta$L) = (0.1, 0.02, 0.03) rlu in a typical geometry, or $\sim$0.43 nm$^{-1}$, full width.

The harmonic phonon frequencies of the high-temperature phase of 1T´-TaTe$_2$ were calculated using density functional perturbation theory (DFPT) \cite{dfpt} with the {\sc Quantum Espresso} package~\cite{0953-8984-21-39-395502,0953-8984-29-46-465901}. The calculations employed the experimental lattice parameters measured at room temperature, with the atomic positions relaxed while preserving the original symmetry of the crystal structure. The Perdew–Burke–Ernzerhof (PBE) approximation~\cite{PBE1996} was used for the exchange-correlation functional. Optimized Norm-Conserving Vanderbilt (ONCV) pseudopotentials~\cite{oncv} were used, including 4$f^{14}$ 5$s^2$ 5$p^6$ 6$s^2$ 5$d^3$ electrons in the valence for Ta, and 4$d^{10}$ 5$s^2$ 5$p^4$ electrons for Te. The plane-wave basis set was truncated at a kinetic energy cutoff of 60 Ry for the wavefunctions. Brillouin zone integrals were performed using a 10$\times$30$\times$8 \textbf{k}-point grid and a Methfessel–Paxton smearing \cite{PhysRevB.40.3616} of 0.0025 Ry.

\section{Acknowledgments}
We thank Matteo Calandra for fruitful discussions and critical reading of the manuscript. S. Mocatti is acknowledged for technical support.  S.K.M. gratefully acknowledges the financial support from the DST-Elettra POC project and the India@DESY collaboration for experiments at Elettra and DESY, respectively. D.S., A. Kar and S.B-C. thank to the MINECO of Spain through the project PID2021-122609NB-C21 and by the European Union Next Generation EU/PRTR-C17.I1, as well as by IKUR Strategy under the collaboration agreement between Ikerbasque Foundation and DIPC on behalf of the Department of Education of the Basque Government. A.K. thanks the Basque government for financial support through the project PIBA-2023-1-0051. C.-Y.L. is supported by the European Research Council (ERC) under the European Union’s Horizon 2020 research and innovation program (grant agreement no. 101020833). V.P, J.C.S and J.P. acknowledge support from Project PID2021-122609NB-C22. J.P. thanks MECD for the financial support received through the "Ayudas para contratos predoctorales para la formación de doctores" grant PRE2019-087338. J.C.S. thanks  the financial support from FPU22/01312. J.D. acknowledges the ERC DELIGHT grant. I.E. acknowledges funding from  the Department of Education, Universities and Research of the Eusko Jaurlaritza, and the University of the Basque Country UPV/EHU (Grant No. IT1527-22); and the Spanish Ministerio de Ciencia e Innovación (Grant No. PID2022-142861NA-I00). A. O. F. thanks the financial support from the Academy of Finland Project No. 369367. We thank Diamond Light Source for access to the ARPES beamline I05 under proposal (si36505) and DESY (Hamburg, Germany), a member of the Helmholtz Association HGF, for the provision of experimental facilities. Parts of this research were carried out at PETRA III using beamline P04. The photoemission spectroscopy instrument at beamline P04 was funded by the German Federal Ministry of Education and Research (BMBF) under the framework program ErUM (projects 05KS7FK2, 05K10FK1, 05K12FK1, 05K13FK1, 05K19FK4 with Kiel University; 05KS7WW1, 05K10WW2, and 05K19WW2 with the University of Würzburg). We acknowledge Spring8 for the provision of IXS beamtime under the proposal 2023A1338. V.P., J.C.S., J.P. thank CESGA (Supercomputing Center of Galicia) for the computing facilities. 



\bibliography{TaTe2_references}

\end{document}


\title{Supplementary information for `First-order phase transition driven by competing charge fluctuations in 1\textit{T}'-TaTe\texorpdfstring{$_2$}{2}}

\author{S. K. Mahatha}
\thanks{These authors contributed equally to this work.}
\affiliation{UGC-DAE Consortium for Scientific Research, University Campus, Khandwa Road, Indore-452001, India}
\affiliation{Ruprecht Haensel Laboratory, Deutsches Elektronen-Synchrotron DESY, 22607 Hamburg, Germany}

\author{A. Kar}
\thanks{These authors contributed equally to this work.}
\affiliation{Donostia International Physics Center (DIPC), Paseo Manuel de Lardizábal, E-20018, San Sebastián, Spain}

\author{J. Corral-Sertal}
\thanks{These authors contributed equally to this work.}
\affiliation{Departamento de Física Aplicada, Universidade de Santiago de Compostela, E-15782 Campus Sur s/n, Santiago de Compostela, Spain}
\affiliation{CiQUS, Centro Singular de Investigacion en Quimica Biolóxica e Materiais Moleculares, Departamento de Quimica-Fisica, Universidade de Santiago de Compostela, Santiago de Compostela, E-15782, Spain}

\author{Josu Diego}
\thanks{These authors contributed equally to this work.}
\affiliation{Department of Physics, University of Trento, Via Sommarive 14, 38123, Povo, Italy}

\author{A. Korshunov}
\affiliation{Donostia International Physics Center (DIPC), Paseo Manuel de Lardizábal. 20018, San Sebastián, Spain}

\author{C.-Y. Lim}
\affiliation{Donostia International Physics Center (DIPC), Paseo Manuel de Lardizábal. 20018, San Sebastián, Spain}

\author{F. K. Diekmann}
\affiliation{Institut f\"{u}r Experimentelle und Angewandte Physik and Ruprecht Haensel Laboratory, Christian-Albrechts-Universit\"{a}t zu Kiel, 24098 Kiel, Germany}

\author{D. Subires}
\affiliation{Donostia International Physics Center (DIPC), Paseo Manuel de Lardizábal, E-20018, San Sebastián, Spain}
\affiliation{University of the Basque Country (UPV/EHU), Basque Country, Bilbao, 48080 Spain}

\author{J. Phillips}
\affiliation{Departamento de Física Aplicada, Universidade de Santiago de Compostela, E-15782 Campus Sur s/n, Santiago de Compostela, Spain}
\affiliation{Instituto de Materiais iMATUS, Universidade de Santiago de Compostela, E-15782 Campus Sur s/n, Santiago de Compostela, Spain} 

\author{T. Kim}
\affiliation{Diamond Light Source Ltd, Harwell Science and Innovation Campus, Didcot, OX11 0DE, United Kingdom}

\author{D. Ishikawa}
\affiliation{Materials Dynamics Laboratory, RIKEN SPring-8 Center, 679-5148 Japan}
\affiliation{Precision Spectroscopy Division, SPring-8 JASRI, 679-5198, Japan}

\author{G. Marini}
\affiliation{Department of Physics, University of Trento, Via Sommarive 14, 38123, Povo, Italy}

\author{I. Vobornik}
\affiliation{CNR-Istituto Officina dei Materiali (CNR-IOM), Strada Statale 14, 34149 Trieste, Italy}

\author{Ion Errea}
\affiliation{Donostia International Physics Center (DIPC), Paseo Manuel de Lardizábal, E-20018, San Sebastián, Spain}
\affiliation{Centro de Física de Materiales (CFM-MPC), CSIC-UPV/EHU, E-20018, San Sebastián, Spain}
\affiliation{Fisika Aplikatua Saila, Gipuzkoako Ingeniaritza Eskola, University of the Basque Country (UPV/EHU), E-20018, San Sebastián, Spain}

\author{S. Rohlf}
\affiliation{Institut f\"{u}r Experimentelle und Angewandte Physik and Ruprecht Haensel Laboratory, Christian-Albrechts-Universit\"{a}t zu Kiel, 24098 Kiel, Germany}

\author{M. Kall\"{a}ne}
\affiliation{Institut f\"{u}r Experimentelle und Angewandte Physik and Ruprecht Haensel Laboratory, Christian-Albrechts-Universit\"{a}t zu Kiel, 24098 Kiel, Germany}

\author{V. Bellini}
\affiliation{Istituto di Nanoscienze, Consiglio Nazionale delle Ricerche, 41125 Modena, Italy}

\author{A.Q.R. Baron}
\affiliation{ Materials Dynamics Laboratory, RIKEN SPring-8 Center, 679-5148 Japan}
\affiliation{Precision Spectroscopy Division, SPring-8 JASRI, 679-5198, Japan}

\author{Adolfo O. Fumega}
\affiliation{Department of Applied Physics, Aalto University, 02150 Espoo, Finland}

\author{A. Bosak}
\affiliation{European Synchrotron Radiation Facility (ESRF), BP 220, F-38043 Grenoble Cedex, France}

\author{V. Pardo}
\email{victor.pardo@usc.es}
\affiliation{Departamento de Física Aplicada, Universidade de Santiago de Compostela, E-15782 Campus Sur s/n, Santiago de Compostela, Spain}
\affiliation{Instituto de Materiais iMATUS, Universidade de Santiago de Compostela, E-15782 Campus Sur s/n, Santiago de Compostela, Spain} 

\author{K. Rossnagel}
\email{rossnagel@physik.uni-kiel.de}
\affiliation{Ruprecht Haensel Laboratory, Deutsches Elektronen-Synchrotron DESY, 22607 Hamburg, Germany}
\affiliation{Institut f\"{u}r Experimentelle und Angewandte Physik and Ruprecht Haensel Laboratory, Christian-Albrechts-Universit\"{a}t zu Kiel, 24098 Kiel, Germany}

\author{S. Blanco-Canosa}
\email{sblanco@dipc.org}
\affiliation{Donostia International Physics Center (DIPC), Paseo Manuel de Lardizábal, E-20018, San Sebastián, Spain}
\affiliation{IKERBASQUE, Basque Foundation for Science, 48013 Bilbao, Spain}

\date{September 2025}

\maketitle
\section{Sample characterization}
\begin{figure}
\includegraphics[width=0.7\linewidth]{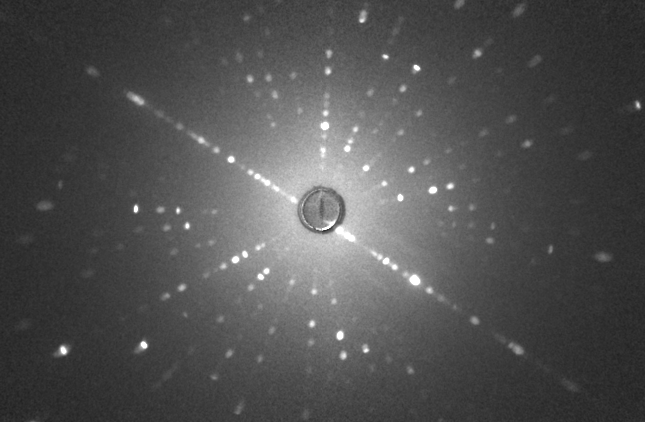}
    \caption{Laue diffraction of 1\textit{T}'-TaTe$_2$ taken at RT.}
    \label{Laue}
\end{figure}

\begin{figure*}
\includegraphics[width=1.0\linewidth]{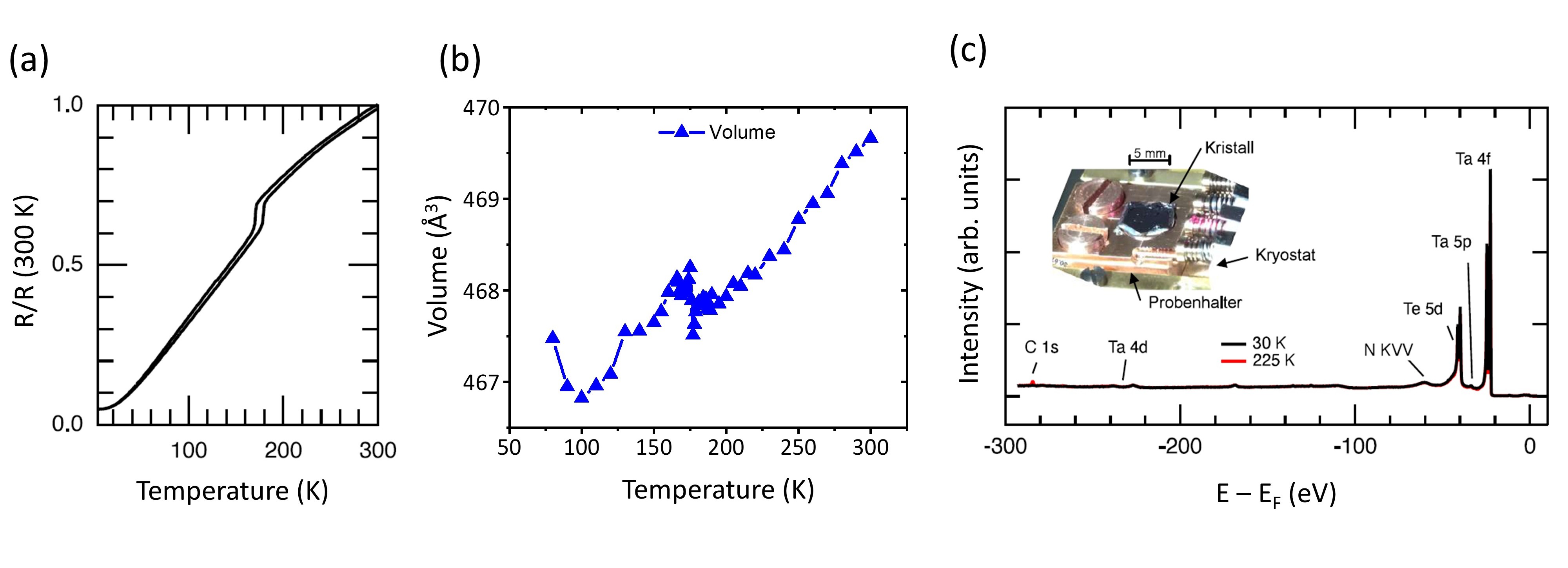}
    \caption{(a) Temperature dependence of the resistivity of 1\textit{T}'-TaTe$_2$, showing a first order transition at T$_\mathrm{CO}$=170 K upon cooling. (b)Temperature dependence of the total volume of 1\textit{T}'-TaTe$_2$, showing a jump at the transition temperature of $\sim$170 K. (c) XPS overview spectrum of the grown 1\textit{T}'-TaTe$_2$ crystals at h$\nu$ = 470 eV, above and below the phase transition temperature (170 K). Inset, photograph of the XPS holder with the 1\textit{T}'-TaTe$_2$ crystal.}
    \label{Resistivity}
\end{figure*}

The high quality of the TaTe$_2$ crystals was first checked with Laue diffraction, Supplementary Fig. \ref{Laue}, and further examined by using a four-contact resistivity measurement on a Quantum Design PPMS and core level photoemission (XPS), Supplementary Fig. \ref{Resistivity}. We observe a residual resistivity ratio (RRR) of  20.4 $\pm$ 0.3, indicating that electron–phonon scattering dominates at room temperature, while low-temperature transport is limited by minimal residual disorder, Supplementary Fig. \ref{Resistivity}(a).  The structural phase transition between the HT and LT phases shows a thermal hysteresis behavior of $\sim$8 K (170.4 $\pm$ 0.3 K (cooling) and 178.8 $\pm$ 0.3 K (heating)). Supplementary Fig.  \ref{Resistivity} (c) shows the result of an XPS measurement (h$\nu$ = 470 eV), showing predominant signatures of tantalum and tellurium in the stoichiometric ratio of TaTe$_2$, and minor impurities of carbon (C 1s) and nitrogen (NKVV Auger and N 1s, respectively).

\section{LEED}
To investigate the quality of the crystalline surface and the evolution of the geometrical structure, we have carried out temperature-dependent LEED measurements are presented in Supplementary Fig. \ref{LEED}. Supplementary Fig. \ref{LEED}(a–b) shows the corresponding LEED diffraction patterns at room temperature (RT) (a) and 30 K (b), taken with a primary electron energy of 80 eV. At RT, the diffraction pattern displays a twofold symmetry of the Bragg diffraction spots. The pattern consists of diffraction reflections from the monoclinic 1$\times$1 phase as well as reflections from a 3$\times$1 superstructure (Supplementary Fig. \ref{LEED}(d)). In samples with lower quality, additional reflections appear between the horizontal diffraction spots. These arise due to spatially averaging LEED measurements capturing contributions from domains rotated by $\pm$60$^\circ$ (Supplementary Fig.  \ref{LEED}(e)). In the low-temperature phase (T $<$ 170 K) (Supplementary Fig. \ref{LEED}(b)), additional diffraction spots emerge, which may indicate a periodic lattice distortion forming a 3$\times$3 superstructure or contributions from $\pm$60$^\circ$ rotated domains. However, certain regions in the diffraction pattern contain reflections that can only be attributed unambiguously to a 3$\times$3 superstructure. One such diffraction spot is highlighted with a blue circle in Supplementary Fig. \ref{LEED}(b). The intensity evolution of this spot as a function of temperature is shown in Supplementary Fig. \ref{LEED}(c). Around 167$\pm$2 K, the intensity increases abruptly and continues to grow during further cooling. This increase in intensity can be explained by the appearance of a diffraction reflection from the 3$\times$3 superstructure and is consistent with the measurement of electrical transport and the temperature-dependent single-crystal XRD measurements. The schematic representation of the decomposition of the LEED diffraction patterns is shown in Supplementary Figs. \ref{LEED}(d–f). Starting from the high-temperature phase (d), the diffraction pattern shows reflections of the monoclinic 1$\times$1 structure (red dot with black outline), which coexists with reflections from the 3$\times$1 structure. Unlike crystals with 1T symmetry (dashed black line), the corresponding surface-projected reciprocal unit cell is significantly more complex. A schematic model consistent with the measured data (red outline) shows a unit cell that is significantly compressed along the horizontal axis, with each center corresponding to a diffraction spot. Contributions from rotated domains are indicated in Supplementary Fig. \ref{LEED}(e) by additional light red dots. In the low-temperature phase, in addition to reflections from the high-temperature phase, those of a 3$\times$3 structure are also observed. Reflections that cannot arise from rotations of the 3$\times$1 structure are marked (blue circle with black outline). The corresponding reciprocal unit cell is represented by the blue hexagon in the Supplementary Fig. \ref{LEED}(f).

\begin{figure*}
    \centering
    \includegraphics[width=0.95\linewidth]{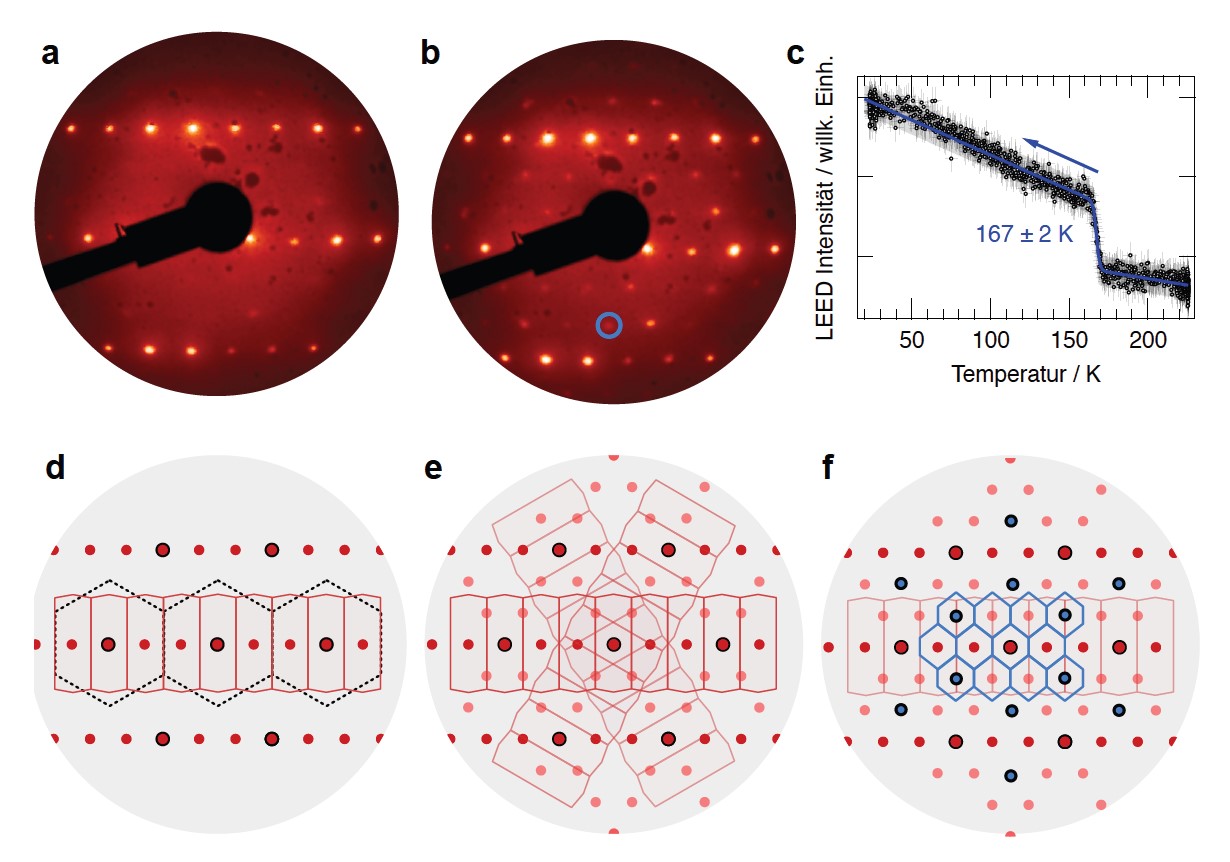}
    \caption{(a) High-temperature and low-temperature structures of 1\textit{T}'-TaTe$_2$ with 3$\times$1 double zigzag chains and 3$\times$3 butterfly structures in reciprocal space. LEED measurement at E$_{kin}$ = 80 eV and (a) 300 K and (b) 30 K, respectively. (c) Temperature dependence of a 3$\times$3 Bragg spot (blue circle in (b)). Schematic LEED model for (d) single-domain 3$\times$1 linear structures (red) with 1$\times$1 diffraction points, (e) contributions from multiple domains ($\pm$60$^\circ$), and (f) 3 $\times$3 superstructure with unique diffraction points (blue).}
    \label{LEED}
\end{figure*}

\section{Soft x-ray ARPES}

\begin{figure*}
\includegraphics[width=1.0\linewidth]{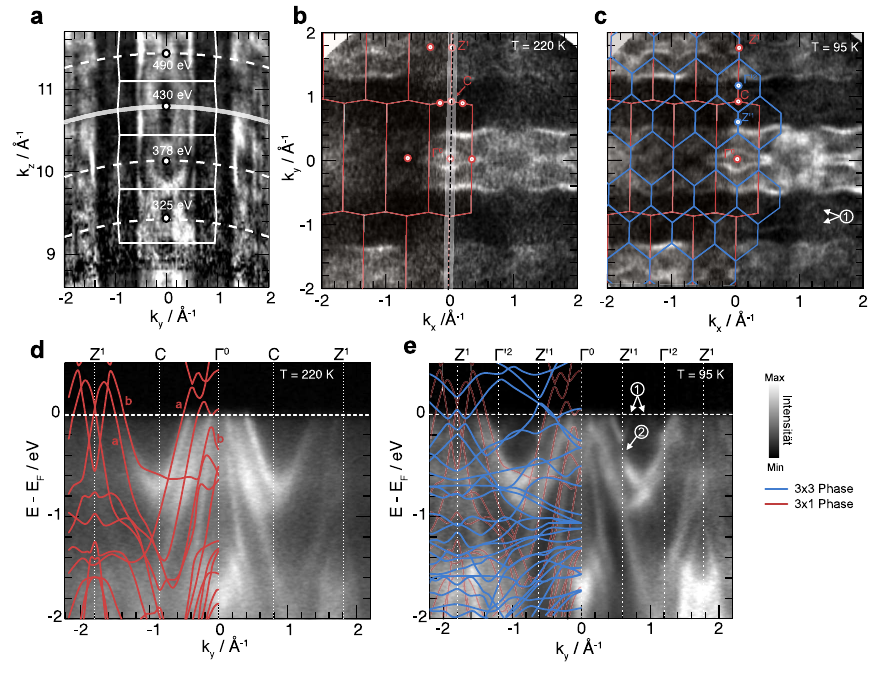}
    \caption{(a) k$_z$ dispersion along $\Gamma^{0}$-Z$^1$, calculated with V$_0$ (inner potential) = 13 eV. The k$_z$ component for ARPES measurements at h$\nu$ = 430 eV is shown in white (bold). Fermi surfaces at (b) 220 K and 95 K, measured at h$\nu$ = 430 eV. The gray area shows the integration interval from which (d) and (e) arise. The experimental band structure along $\Gamma^{0}$-Z$^1$ for (d) 220 K and (e) 95 K compared to DFT calculations of the HT phase (red) and LT phase (blue).}
    \label{soft-ARPES}
\end{figure*}

To understand and investigate the phase transition behavior over a larger reciprocal space with multiple Brillouin zones, the valence band mapping and Fermi surfaces were obtained at 430 eV photon energy (Supplementary Fig. \ref{soft-ARPES}). The photon energy dependent spectra exhibit a periodic modulation of the band structure in the k$_z$-k$_y$ plane (Supplementary Fig. \ref{soft-ARPES}(a)), and are shaped by the one-dimensional character of the 3$\times$1 structure, showing only weakly dispersive features in the region around k$_y$ = $\pm$0.4 \AA$^{-1}$. Within a photon energy series, the valence bands appeared with the highest intensity at 430 eV photon energy, and the projected Brillouin zone (BZ) passes through the center of the 15$^{th}$ BZ (for V$_0$ = 13 eV). Changing photon polarization and experimental geometry leads to a significant change in the intensity variation of the photoemitted spectra, indicating a strong dependence on matrix elements. Therefore, the analysis focuses primarily on energy changes during the phase transition. Supplementary Fig. \ref{soft-ARPES}(b) and (c) represent the Fermi surface mapping at high (220 K) and low (95 K) temperatures, respectively. The investigated k$_\parallel$ region is characterized by three horizontal stripes with high photoemission intensity, and the center of each stripe represents the vertical center of BZ. For the high-temperature phase, the high-symmetry point, $\Gamma^{\circ}$, is characterized by an electron pocket. This electron pocket is flanked by two wave-like, horizontally running features that mark the edges of the intensity stripes (Supplementary Fig. \ref{soft-ARPES}(b)). The extrema of these modulations correspond to the centers or edges of the BZ. Upon cooling the sample well below 170 K, thermal broadening of the bands decreases, and individual bands become better resolved. Supplementary Fig. \ref{soft-ARPES}(c) shows the Fermi surface at 95 K. The structural features of the Fermi surface known from the HT phase largely remain, although additional fine features appear in the less intense regions between the horizontally extended, chain-like structures. These arise from the 3$\times$3 modulation and the possible back folding of valence bands along $\Gamma^{\circ}-Z^{1}$.
Below the transition temperature of approximately 170 K, the calculated band structure changes significantly. In the measured valence band structure (Supplementary Fig. \ref{soft-ARPES} (e)) at 95 K along $\Gamma^{\circ}-Z^{1}$, the band structure observed from the HT phase still dominates. However, changes in band dispersion at 1/3 $\Gamma^{\circ}-Z^{1}$ ($Z^{'1}$) and 2/3 $\Gamma^{\circ}-Z^{1}$ ($\Gamma^{'2}$) are attributed to back-folding from the 3$\times$3 superstructure. From the previously observed dispersive V-shaped band, two additional V-shaped bands now emerge, with minima at $Z^{'1}$ and $\Gamma^{'2}$
(marked as 1 in Supplementary Fig. \ref{soft-ARPES}(e)). These orbitals are primarily associated with Ta atoms and generate a band gap of approximately 100 meV below the additional V-shaped bands at $Z^{'1}$ and $\Gamma^{'2}$ (marked as 2 in Fig. \ref{soft-ARPES}(e)). In the region around the high-symmetry point $\Gamma^{\circ}$, the calculated band structure does not match the ARPES data at low temperature. While DFT calculations predict strong changes in dispersion compared to the HT phase in the Ta and Te states, the measured band structure more closely resembles the HT phase and is characterized by highly dispersive bands. The discrepancy between measured and calculated band structures suggests that the low-temperature (LT) phase does not consist solely of a 3$\times$3 structure, but rather emerges from a combination of both 3$\times$3 and 3$\times$1 structures.

The current results from Ta 4\textit{f} core-level electrons and ARPES measurements during the structural phase transition show no evidence of an energy gap near the Fermi level or a CDW-induced splitting. Instead, TaTe$_2$ undergoes a reconstruction of the band structure in the low-energy range as the system passes through the structural phase transition, indicating a strong electron–phonon interaction.

\section{DFT}
\begin{figure*}
\includegraphics[width=1.0\linewidth]{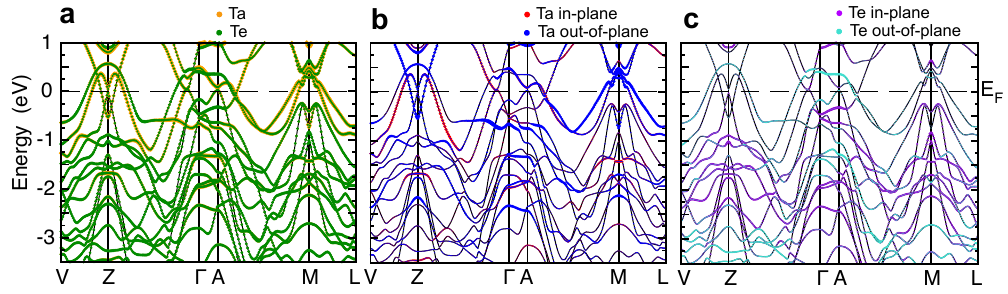}
    \caption{Atom and orbital resolved DFT calculations of the 1\textit{T}'-TaTe$_2$. Bands at the Fermi level are mostly derived from the in- and out-of-plane Ta orbitals.}
    \label{DFT}
\end{figure*}

From DFT, we observe that the valence bands are mainly dominated by the Ta 5\textit{d} states, which are separated from the Te 5\textit{p} states, located below 1 eV binding energy, Supplementary Fig. \ref{DFT}. In the region around the high-symmetry point $\Gamma$, we can identify approximately four dispersive V-shaped bands crossing the Fermi level, which are symmetric about the C point. Some Te 5\textit{p} states overlap with the Ta 5\textit{d} states at around 0.6 eV binding energy and exhibit a strongly dispersive V-shaped dispersion. The band minimum is located at the high-symmetry point C (Supplementary Fig. \ref{soft-ARPES}(e)). The band crossings at the Fermi level give rise to the prominent outer wave-like structures observed in the Fermi surface of the zigzag chains.

\begin{figure*}
\includegraphics[width=0.8\linewidth]{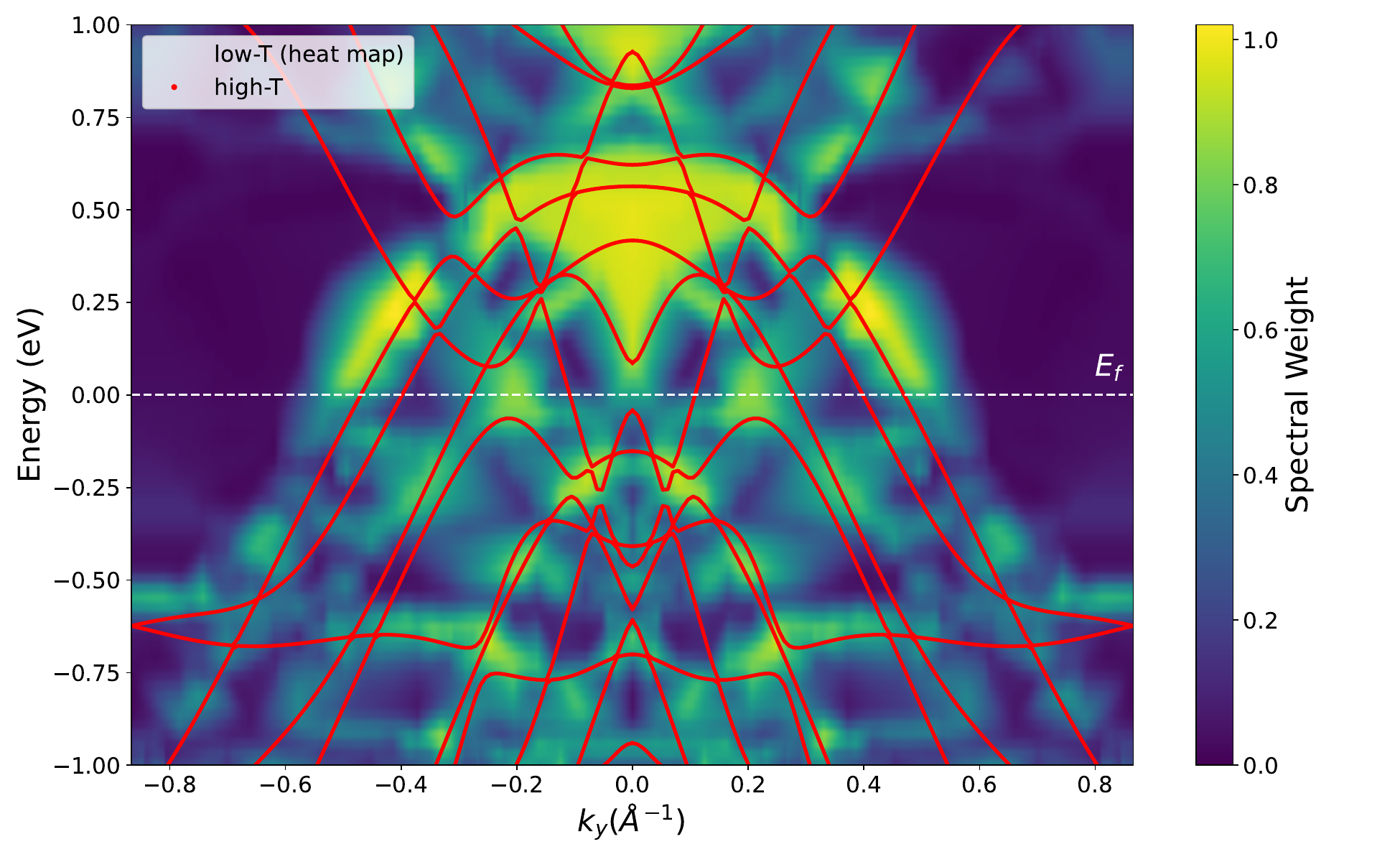}
    \caption{DFT calculated band structure of 1\textit{T}'-TaTe$_2$ : A comparison of high (3$\times$1) and low (3$\times$3) temperature.}
    \label{unfold}
\end{figure*}

Supplementary Fig. \ref{unfold} shows together a representation of the band structure for both the high temperature and the low temperature structures in the same Brillouin zone. The low-temperature supercell band structure was unfolded to the high-temperature cell (smaller in real space, larger in k-space). The unfolded band structure is presented as a heat map where the intensity of the map represents the spectral weight at that particular energy and
\textit{k}-vector. The heat map corresponding to the low-temperature phase shows two high-intensity crossings at the Fermi level, one at about 0.25 \AA$^{-1}$ and the other at 0.55 \AA$^{-1}$. The latter has a band of the high-temperature phase coinciding on top of it, but the former is displaced towards $\Gamma$ with respect to the high-temperature phase. This agrees qualitatively with the results shown in Fig. 3(F) in the main text (which discusses crossings at about 0.25 \AA$^{-1}$ and 0.45 \AA$^{-1}$). This one is slightly off-predicted by DFT, although the temperature trend is correct.

\section{Identification of the nesting vectors}

\begin{figure*}
\includegraphics[width=0.8\linewidth]{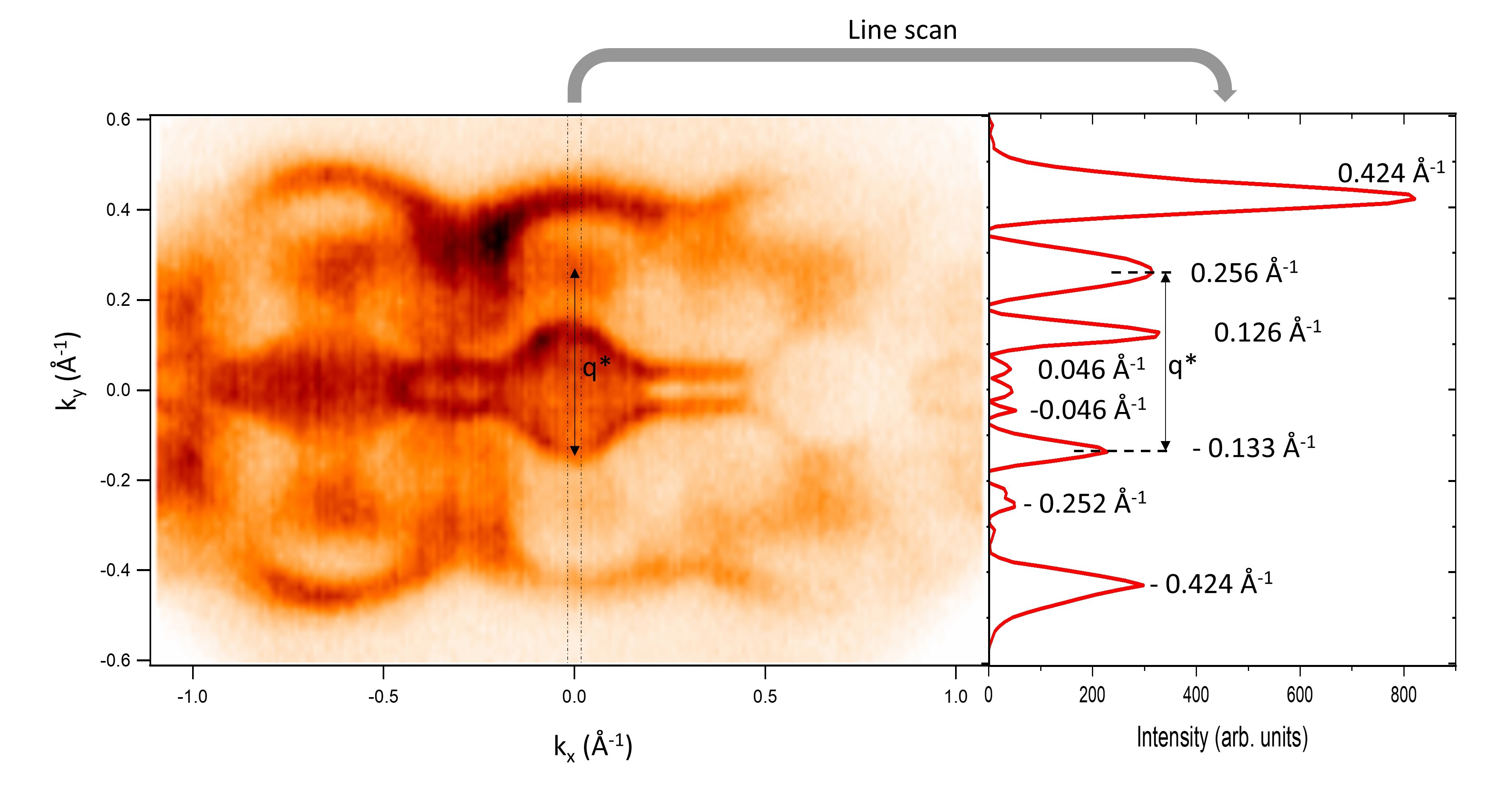}
    \caption{Fermi surface (left panel) of 1\textit{T}'-TaTe$_2$ obtained at 20 K, 65 eV photon energy with LH polarized light. The figure exhibits the reciprocal space scaling of the nesting vector (q$^*$). The right panel shows the baseline corrected line scan spectra taken along $k_y$ at $k_x$ = 0.}
    \label{ARPES}
\end{figure*}

From the high resolution ARPES spectra taken with 65 eV at 20 K, Fig. 3(D) in the main text, we can infer several nesting vectors from parallel segments of the Fermi surface. Supplementary Fig. \ref{ARPES} displays the Fermi surface and the intensity profile of the bands perpendicular to the chain direction. We see that, among all the possible nesting vectors, the states connected by the Fermi wavevectors at $\sim-0.14$ and $\sim0.26$ characterize nearly parallel segments. This vector distance nearly matches the q$^*$=(0\ $\frac{1}{4}+\delta$\ 0) wavevector, hence linking the Fermi surface with the charge fluctuations observed in DS. 

\section{Electronic susceptibility and nesting function}

To understand the role of nesting in the formation of the CDW, we analyzed the non-interacting electronic susceptibility. The nesting function $\zeta(\mathbf{q})$, connected to the imaginary part of the susceptibility, identifies wavevectors $\mathbf{q}$ connecting multiple regions of the Fermi surface:

\begin{equation}
    \zeta(\mathbf{q}) = \frac{1}{N_\mathbf{k}} \sum_{nn'} \sum_{\mathbf{k}}^{1BZ} \delta(\epsilon_{n\mathbf{k}}) \delta(\epsilon_{n'\mathbf{k}+\mathbf{q}}) \,.
 \tag{Supplementary Eq. 1}
\end{equation}

While $\zeta(\mathbf{q})$ provides insight into the topology of the Fermi surface, it is the real part of the susceptibility, $\chi_0(\mathbf{q})$, that must diverge for nesting to be the driving mechanism behind the transition:
\begin{equation}
    \chi_0(\mathbf{q}) = \frac{1}{N_\mathbf{k}} \mathcal{P} \sum_{nn'} \sum_{\mathbf{k}}^{1BZ} \frac{f_{n\mathbf{k}} - f_{n'\mathbf{k}+\mathbf{q}}}{\epsilon_{n\mathbf{k}} - \epsilon_{n'\mathbf{k}+\mathbf{q}}} \,,
     \tag{Supplementary Eq. 2}
\end{equation}

where $\mathcal{P}$ denotes the principal value. This effect of nesting is also present in the renormalization of harmonic phonon frequencies due to electron-phonon interaction in DFPT calculations, which arises from the real part of the static limit of the electron-phonon self-energy:

\begin{equation}
    \Pi_{\mu}(\mathbf{q}, \omega=0) = \frac{1}{N_\mathbf{k}} \sum_{nn'} \sum_{\mathbf{k}}^{1BZ} \left| g^{\mu}_{n'\mathbf{k}+\mathbf{q}, n\mathbf{k}} \right|^2 \frac{f_{n\mathbf{k}} - f_{n'\mathbf{k}+\mathbf{q}}}{\epsilon_{n\mathbf{k}} - \epsilon_{n'\mathbf{k}+\mathbf{q}}} \,,
     \tag{Supplementary Eq. 3}
\end{equation}

where the electron-phonon matrix elements $g^{\mu}_{n'\mathbf{k}+\mathbf{q}, n\mathbf{k}}$ explicitly appear. We aim to identify whether nesting or the electron-phonon matrix elements play the primary role in triggering the instability observed in the harmonic phonons.

$\zeta(\mathbf{q})$ and $\chi_0(\mathbf{q})$ were calculated in the high-temperature phase of 1\textit{T}'-TaTe$_2$ with the EPIq code \cite{MARINI2024108950}, using maximally localized Wannier Functions (MLWF) for entangled bands \cite{MLWF, MLWF2} as implemented in the Wannier90 code \cite{Wan90}. We obtained MLWF by using 60 Wannier functions initialized from random projections and a 2$\times$9$\times$5 \textbf{k}-point grid. $\zeta(\mathbf{q})$ and $\chi_0(\mathbf{q})$ where then calculated with 50$\times$150$\times$100 \textbf{k}-point grids. A smearing of 0.01 eV was applied in the computation of the nesting function.

\begin{figure*}
\includegraphics[width=1.0\linewidth]{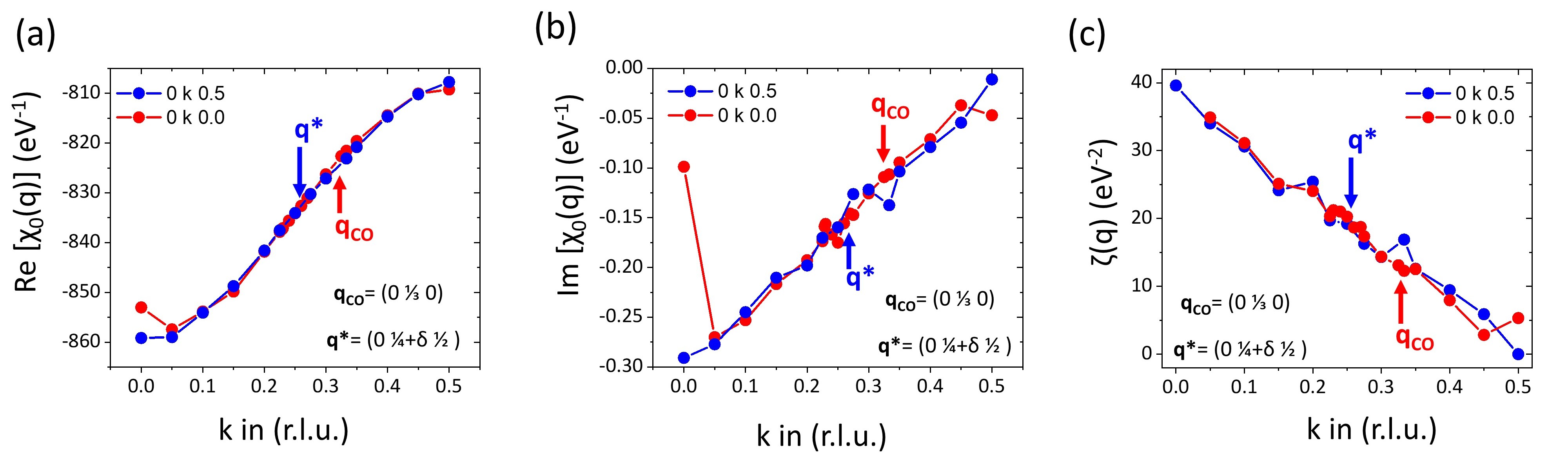}
    \caption{Calculated electronic susceptibility along the k-direction. (a) Real and (b) imaginary parts of the electronic susceptibility. (c) Nesting function.}
    \label{nesting}
\end{figure*}

The results of $\zeta(\mathbf{q})$ and $\chi_0(\mathbf{q})$ are presented in Supplementary Fig. \ref{nesting}. The calculations do not show any peak at both q$^*$ and q$_\mathrm{CO}$, although the nesting values are anomalously large in the k-space.  

\section{Phonon calculations}

\begin{figure}
    \centering
    \includegraphics[width=0.5\linewidth]{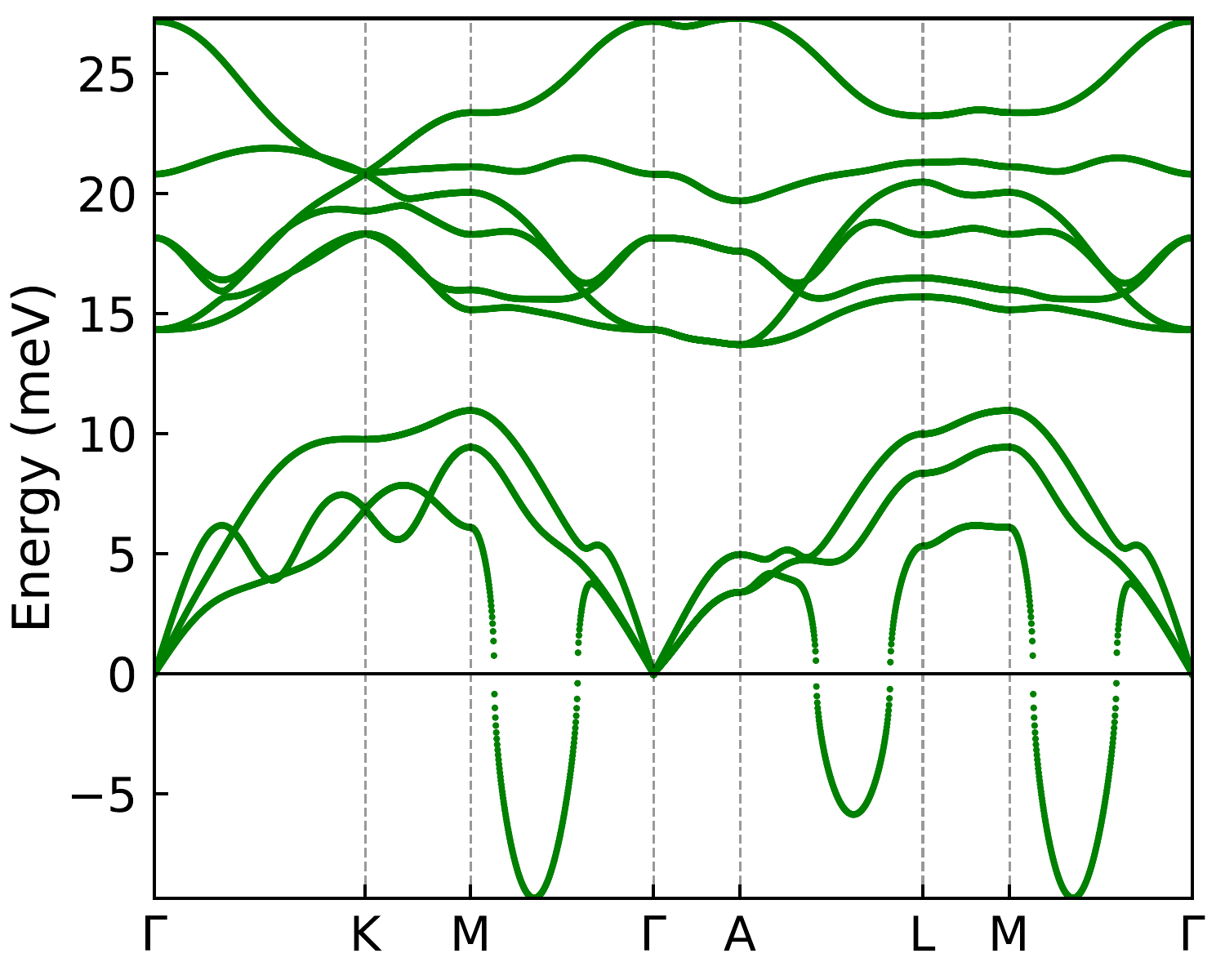}
    \caption{Phonon band dispersion in the hypothetical normal state configuration 1T-TaTe$_2$.}
    \label{phon_NS}
\end{figure}

Phonon dispersion were calculated both with QUANTUM ESPRESSO (main text) and VASP (supplementary information) codes.  First, we compute the lattice dynamics of the hypothetical 1T-phase of TaTe$_2$ (similar to the 1T-VSe$_2$). Supplementary Fig. \ref{phon_NS} displays an imaginary frequency of the low-energy acoustic mode at ($\frac{1}{3}$\ 0\ 0), in agreement with the experimentally observed 1\textit{T}'-TaTe$_2$.

\begin{figure*}
\includegraphics[width=0.8\linewidth]{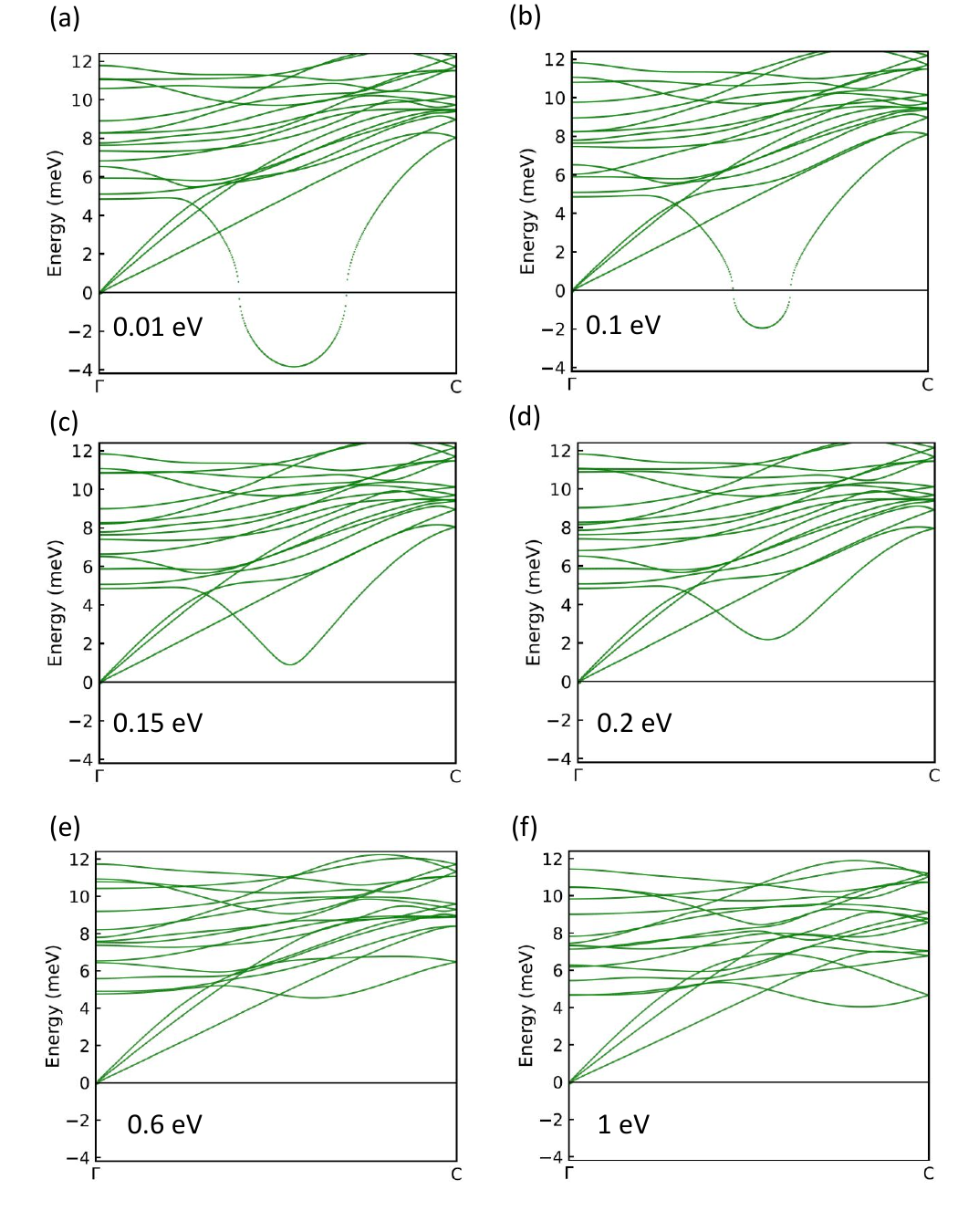}
    \caption{Phonon dispersion showing the imaginary mode at (0\ $\sim\frac{1}{4}$\ 0) calculated with (a) 0.01 eV, (b) 0.1 eV, (c) 0.15 eV, (d) 0.2 eV, (e) 0.6 eV and (f) 1 eV smearing. Calculations were carried out with the VASP code.}
    \label{smearing}
\end{figure*}

Supplementary Fig. \ref{smearing} shows the VASP calculations with different smearing values that mimic the effect of the electronic temperature. Furthermore, we see that the imaginary mode at q$^*$ becomes stable with 0.15 eV smearing, hence suggesting the electron phonon interaction contributes to the phonon instability.
 
\section{Inelastic x-ray Scattering}

\begin{figure*}
\includegraphics[width=0.75\linewidth]{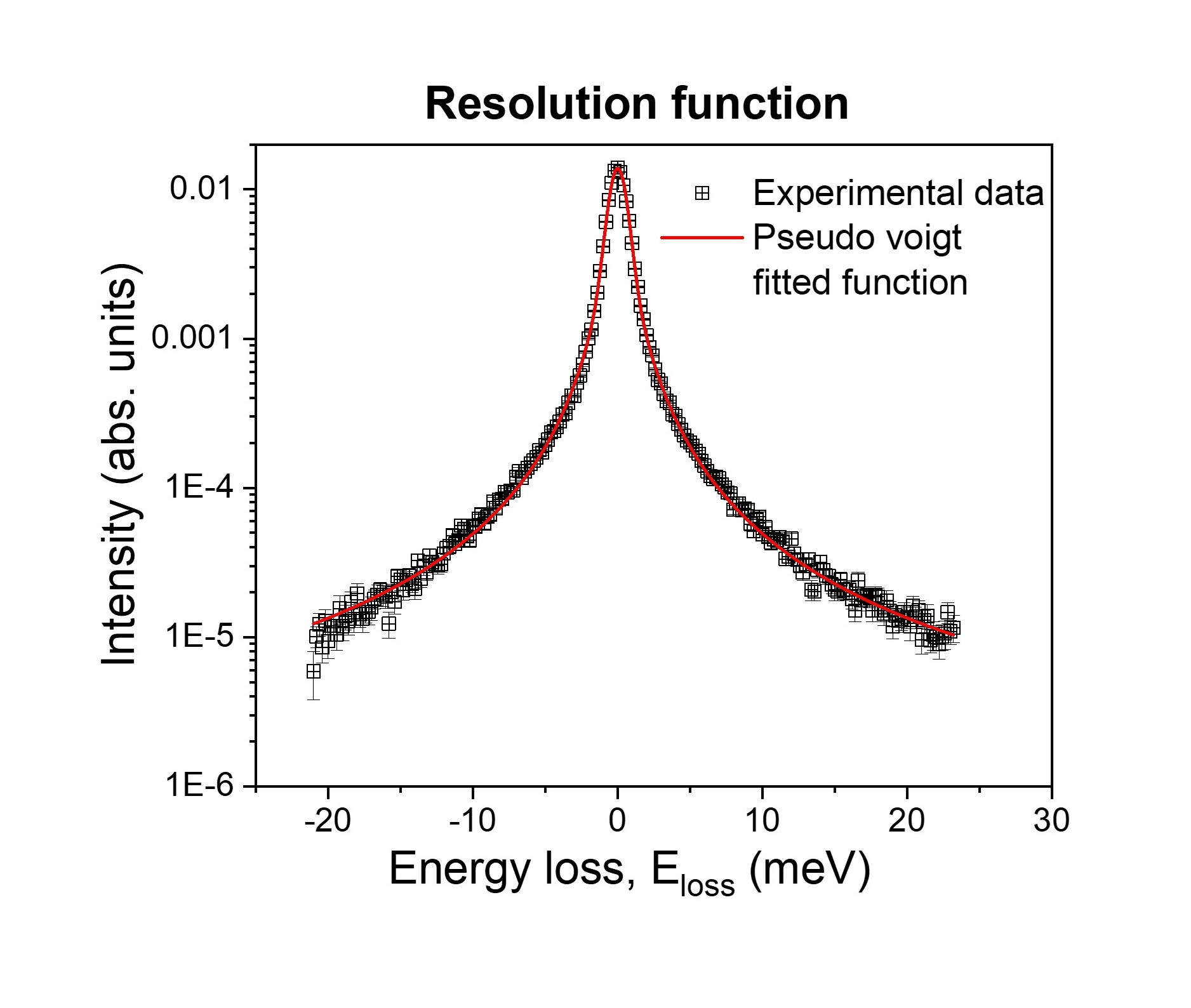}
    \caption{Fitting of Inelastic x-ray Scattering spectrometer resolution function with a pseudo-Voigt profile.}
    \label{IXS_resolution}
\end{figure*}

In this section, we will describe the fitting procedure we used for the High-resolution inelastic x-ray scattering measurements carried out at BL43LXU \cite{Baron2019,Baron2020} of SPring-8. Inelastic x-ray scattering (IXS) measurements were made as a function of temperature and momentum with energy and momentum resolution of $\sim$1.4 meV and 0.4 nm$^{-1}$, respectively, with incident beam energy of 21.75 keV. The horizontally scattered beam was analyzed by a silicon analyzer, Si (11, 11, 11). While a more precise scheme is available \cite{Ishikawa_2021}, we chose to model the resolution function as the measured response of Tempax glass near the structure factor maximum – a mostly elastic scatterer -  as a pseudo-Voigt profile, Supplementary Fig. \ref{IXS_resolution}.

\begin{equation}
y = y_0 + \frac{2A}{\pi} \left( 
\mu \cdot \frac{w_L}{4(x-x_c)^2 + w_L^2} + 
(1-\mu) \cdot \frac{\sqrt{4 \ln 2}}{\sqrt{\pi} w_G} \exp\left( 
-\frac{4 \ln 2}{w_G^2} (x - x_c)^2 
\right) 
\right)
 \tag{Supplementary Eq. 4}
\end{equation}
Whereas $\mu$ is the Lorentz factor, $w_L$ and $w_G$ are the Lorentzian and Gaussian linewidths, respectively. We obtain the following from the fitting: $\mu$ =0.61966, $w_L$ = 1.75892, and $w_G$ = 1.30555.

\textbf{Fitting of IXS spectra:}
The fitting of the IXS spectra at a given temperature is performed using damped harmonic oscillators (DHO) for phonons, convoluted with the experimental resolution. The dynamic structure factor \( S(\mathrm{Q}, \omega) \) is given by:

\begin{equation}
S(\mathrm{Q}, \omega) = \frac{[n(\omega) + 1]\, Z(\mathrm{Q})\, 4\omega\, \frac{\Gamma_{\mathrm{q}}}{\pi}}{[(\omega - \omega_{\mathrm{q}})^2 + \Gamma_{\mathrm{q}}^2][(\omega + \omega_{\mathrm{q}})^2 + \Gamma_{\mathrm{q}}^2]} \tag{Supplementary Eq. 5}
\end{equation}

where \( Z(\mathrm{Q}) = \exp(-2W_{\mathrm{Q}})\, |\mathrm{Q} \cdot \mathrm{e}|^2 / 2M \), with the exponential term representing the Debye-Waller factor, \( \mathrm{e} \) being the polarization vector, and \( M \) the mass of the atom. Representative IXS analysis is shown in Supplementary Figs. \ref{IXS_1}, \ref{IXS}, \ref{IXS_2} and \ref{IXS_3}.

\begin{figure*}
\includegraphics[width=1.0\linewidth]{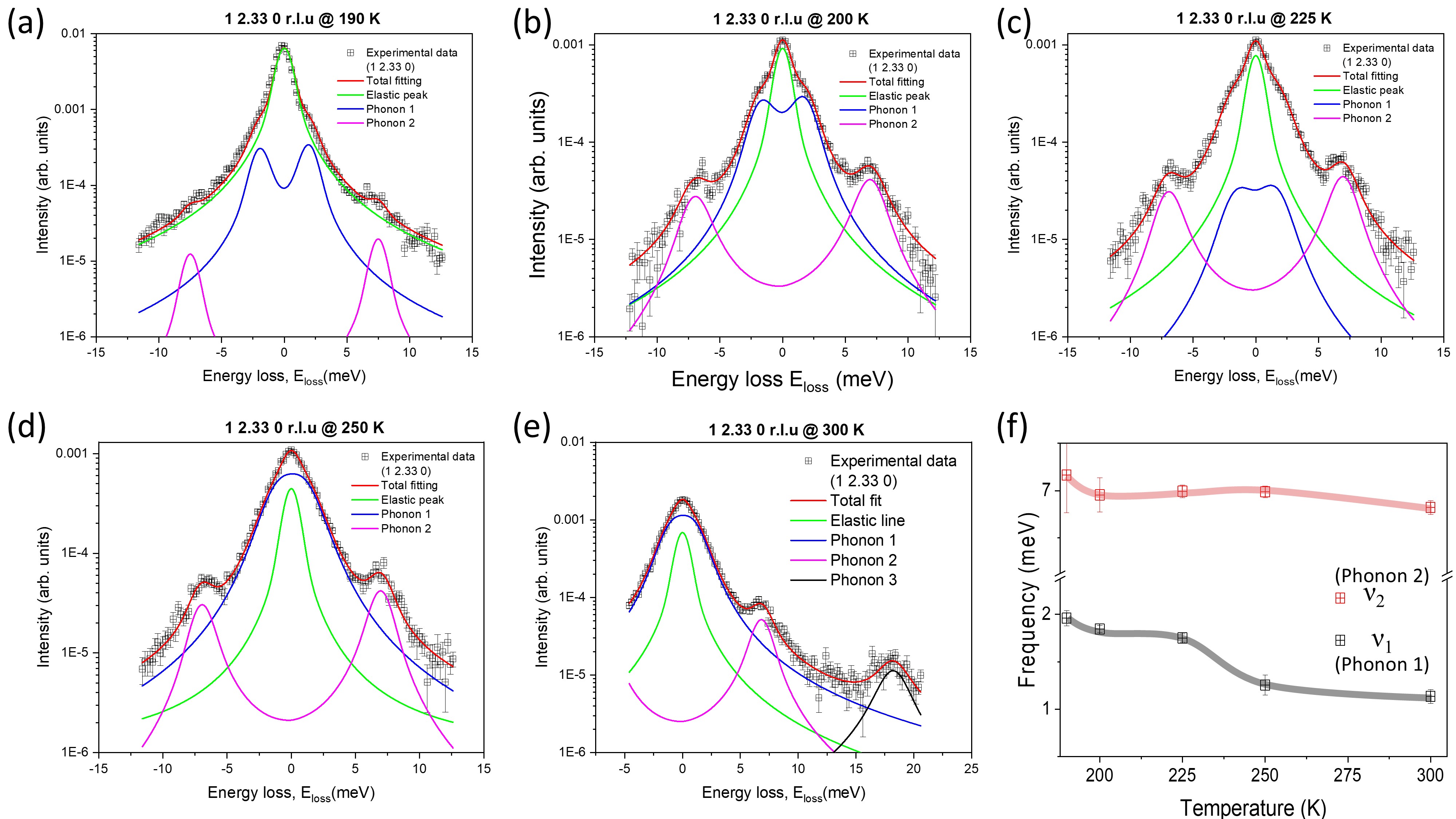}
  \caption{
  \textbf{Fitting of IXS spectra at (1, 2.33, 0) r.l.u.} 
        (a)–(e) Each panel shows data measured at a different temperature. The total fit to the experimental spectra is shown as a red line. The elastic line is plotted in green, while the two phonon modes—phonon 1 (\( \nu_1 \)) and phonon 2 (\( \nu_2 \))—are fitted with blue and pink lines, respectively. 
        (f) Temperature dependence of the phonon frequencies.}
    \label{IXS_1}
\end{figure*}

\begin{figure*}
\includegraphics[width=1.0\linewidth]{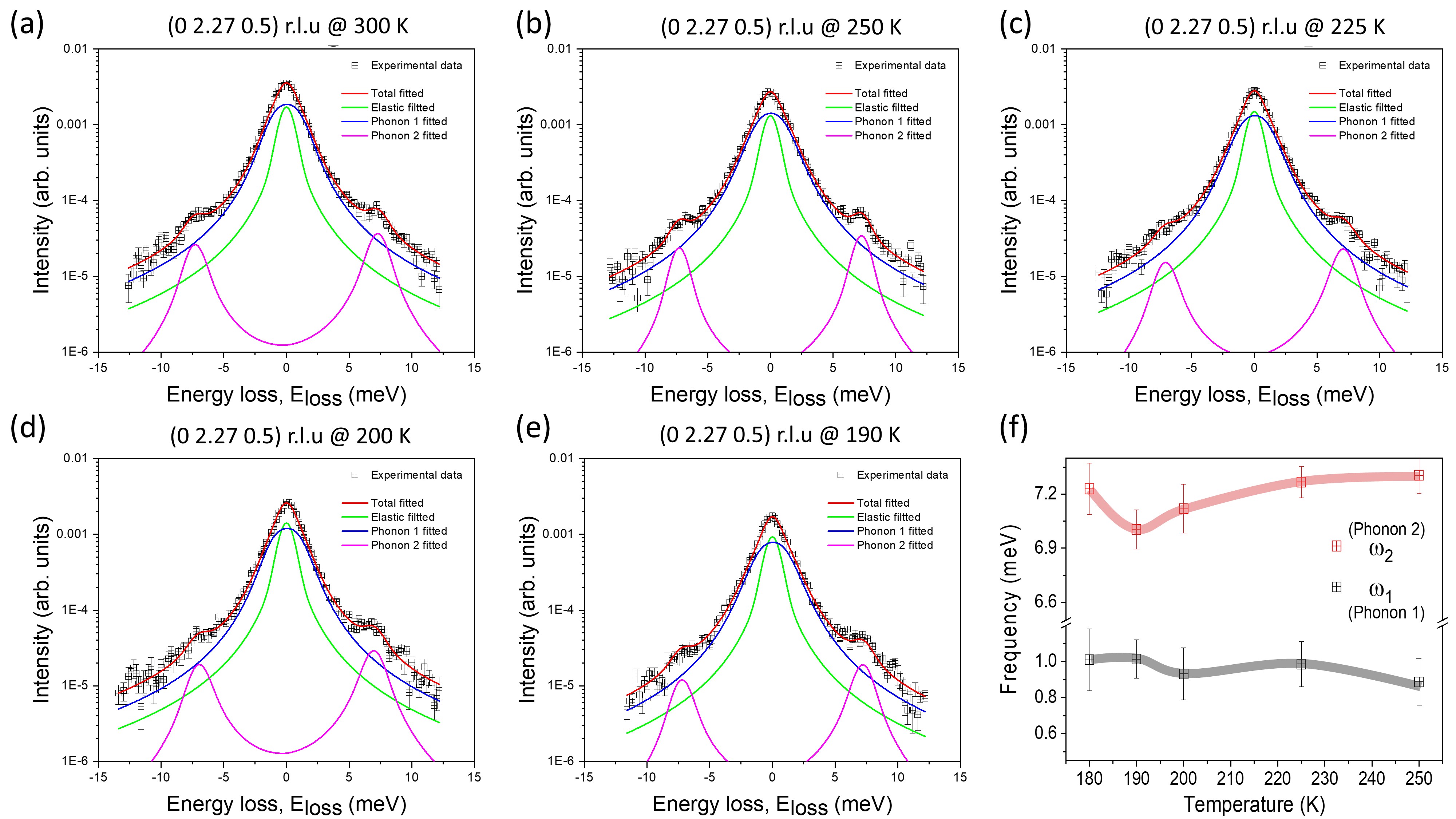}
    \caption{\textbf{Fitting of IXS spectra at (0, 2.27, 0) r.l.u.} 
        (a)–(e) Each panel shows data measured at a different temperature. The total fit to the experimental spectra is shown as a red line. The elastic line is plotted in green, while the two phonon modes—phonon 1 (\( \omega_1 \)) and phonon 2 (\( \omega_2 \))—are fitted with blue and pink lines, respectively. 
        (f) Temperature dependence of the phonon frequencies at (0, 2.27, 0.0) r.l.u. momentum.}
    \label{IXS}
\end{figure*}

\begin{figure*}
\includegraphics[width=1.0\linewidth]{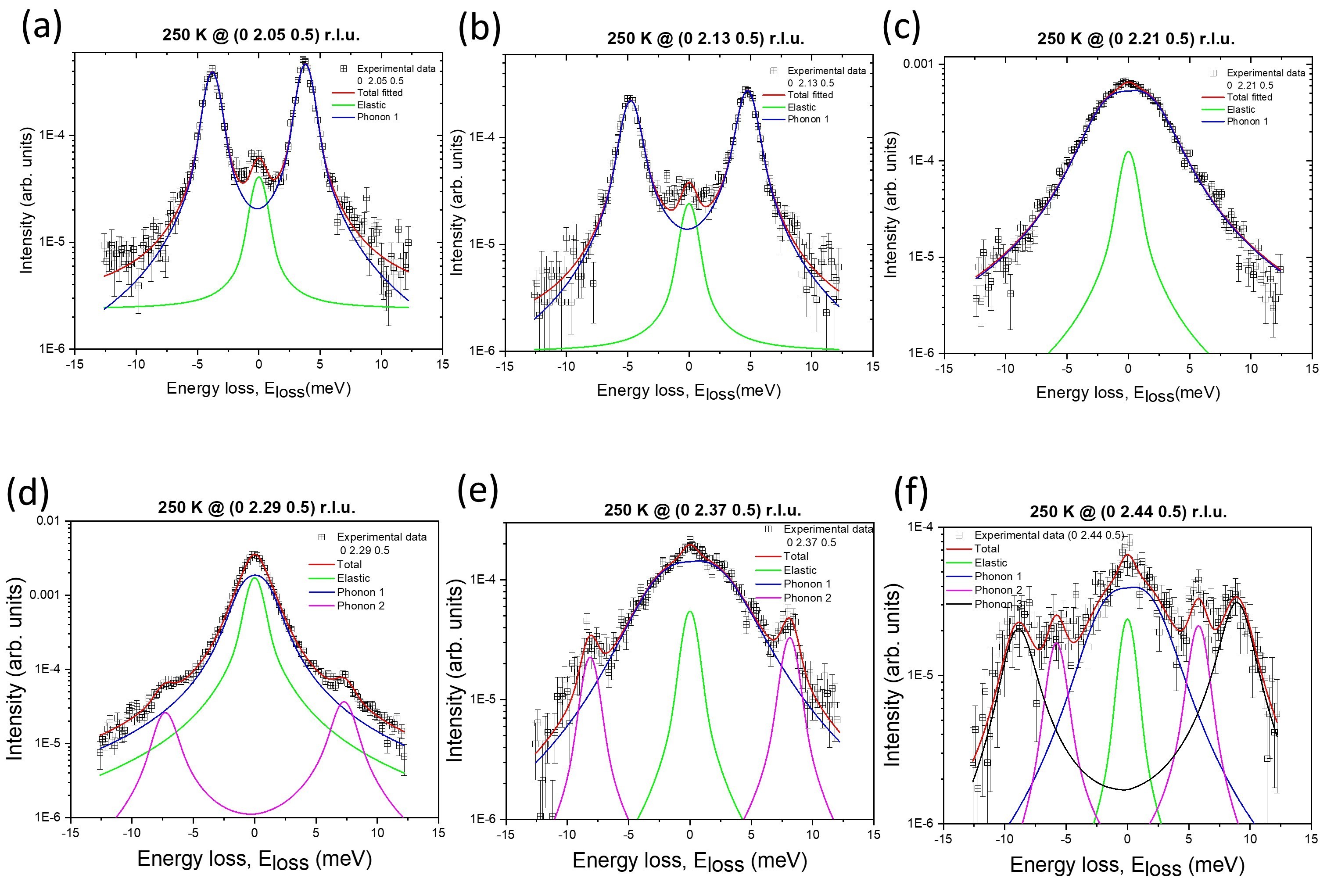}
    \caption{\textbf{Fitting of IXS spectra at 250 K along (0, 2+k, 0.5) r.l.u. direction.} (a)-(f) Each panel corresponds to different momentum along  k= 2+k momentum direction. The total fit to the experimental spectra is shown as a red line. The elastic line is plotted in green, while the two phonon modes—phonon 1  and phonon 2 —are fitted with blue and pink lines, respectively. } 
    \label{IXS_2}
\end{figure*}

\begin{figure*}
\includegraphics[width=1.0\linewidth]{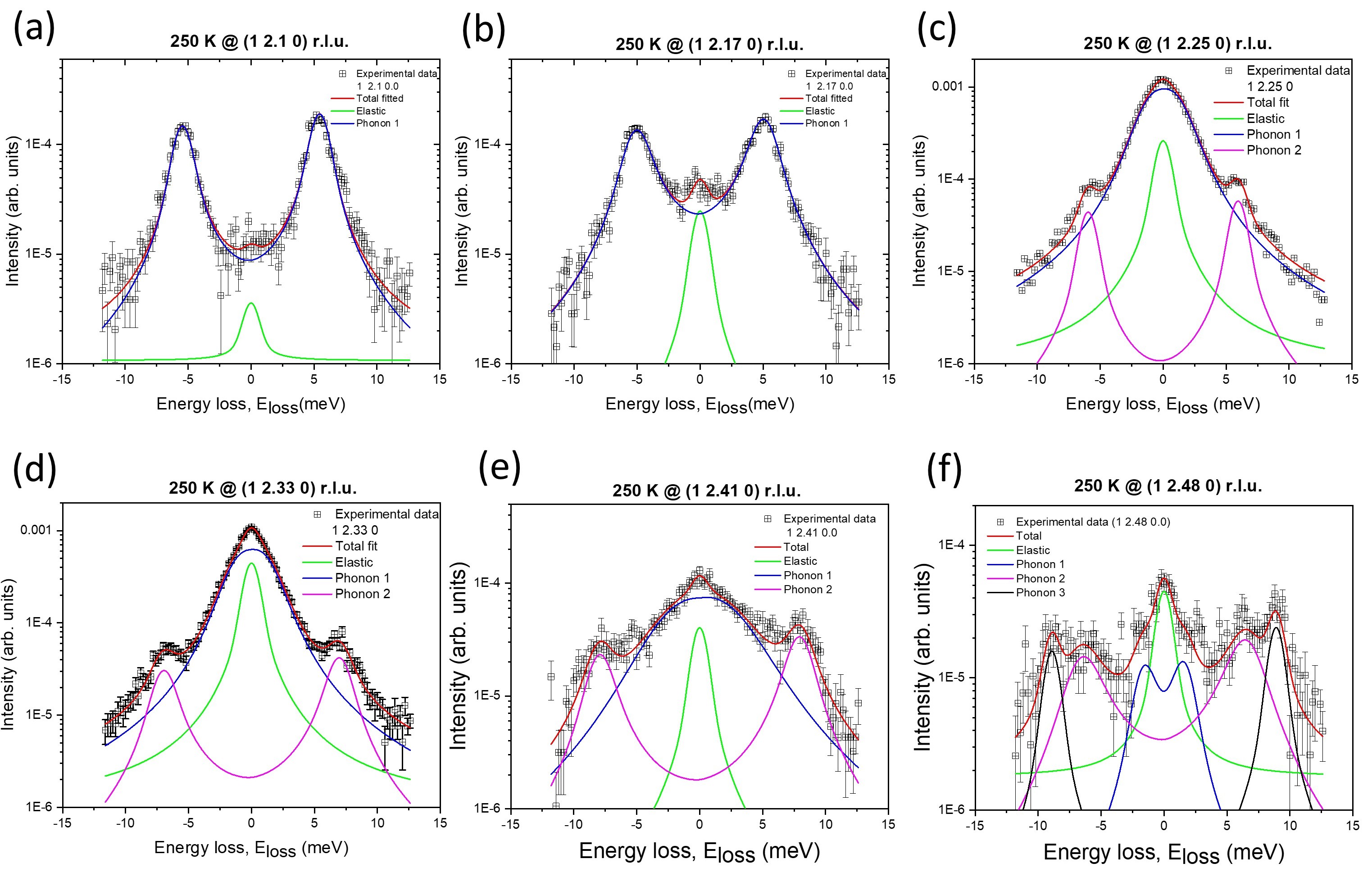}
    \caption{\textbf{Fitting of IXS spectra at 250 K along (1, 2+k, 0.0) r.l.u. direction.} (a)-(f) Each panel corresponds to different momentum along  k= 2+k momentum direction. The total fit to the experimental spectra is shown as a red line. The elastic line is plotted in green, while the two phonon modes—phonon 1  and phonon 2 —are fitted with blue and pink lines, respectively.}
    \label{IXS_3}
\end{figure*}

\section{Spectral intensity of the low energy modes}
\begin{figure*}
\includegraphics[width=1.0\linewidth]{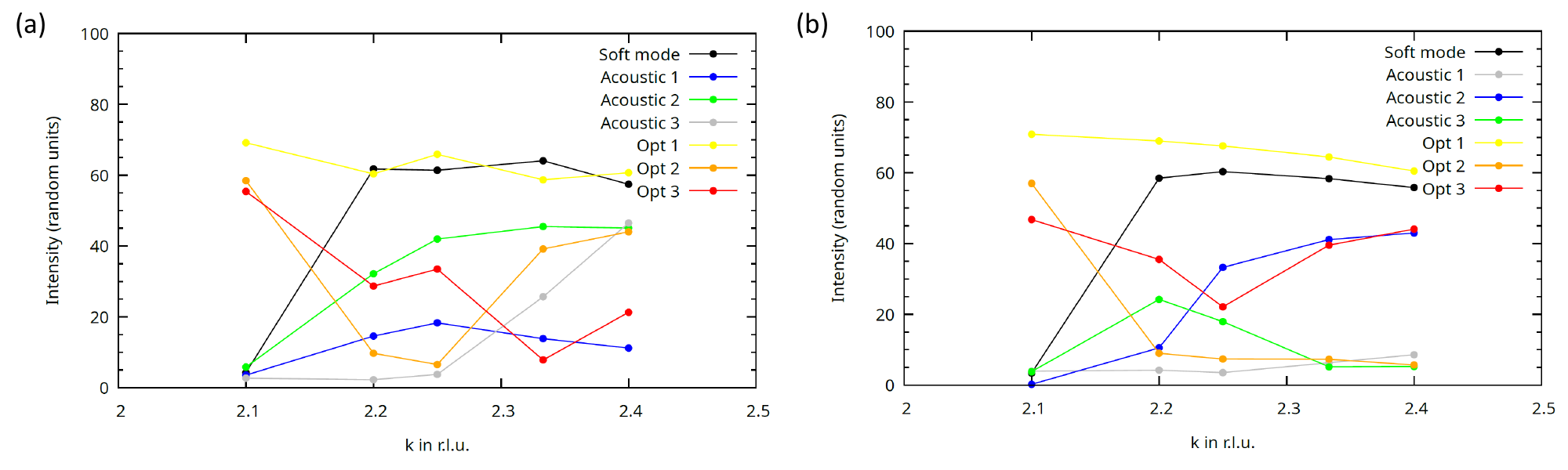}
    \caption{Momentum dependence of the IXS spectral intensity for the low energy modes along the (a) (0\ 2+K\ $\frac{1}{2}$) and (b) (1\ 2+K\ 0) paths.}
    \label{phonon-int}
\end{figure*}

Supplementary Fig. \ref{phonon-int} shows the momentum dependence of the IXS intensity of the low energy modes along the (0\ 2+K\ $\frac{1}{2}$) and the (1\ 2+K\ 0) paths. In both panels (a) and (b), in the momentum range K$<$0.2, the spectral weight of IXS is dominated by the optical modes $\omega_2$ and $\nu_2$. Between 0.2$<$K$<$0.4, the IXS intensity contains spectral weight from $\omega_1$ and $\nu_1$ (soft modes) and $\omega_2$ and $\nu_2$. Nevertheless, the strong renormalization of $\omega_1$ and $\nu_1$ (1-2 meV) due to the EPI allows us to distinguish $\omega_1$ from $\nu_1$ and $\omega_2$ from $\nu_2$. 

\section{Ginzburg-Landau model}

Here, we show a minimal Ginzburg-Landau model describing the competition between two charge density wave (CDW) phases, q$^{*}$ and q$_\mathrm{CO}$, observed in the layered van der Waals material 1\textit{T}'-TaTe$_2$. The model captures the emergence of the q$^{*}$ phase near 250~K, followed by a first-order transition to the $q_\mathrm{CO}$ CDW phase at approximately 180~K. This is achieved by including a biquadratic competition between the two complex scalar order parameters. To describe this, we construct a Ginzburg-Landau free energy with two competing order parameters.

\subsection{Model}
We define two complex scalar order parameters:
\begin{itemize}
    \item $\psi_\mathrm{CO}$: amplitude of the $q_\mathrm{CO}$ CDW
    \item $\psi^{*}$: amplitude of the $q^{*}$ CDW
\end{itemize}

The total free energy density is:
\begin{align}
F &= a_1(T)|\psi_{CO}|^2 + \frac{b_1}{2}|\psi_{CO}|^4 
  + a_2(T)|\psi^{*}|^2 + \frac{b_2}{2}|\psi^{*}|^4 + \gamma |\psi_{CO}|^2 |\psi^{*}|^2
\end{align}

Temperature dependence is modeled via:
\begin{align}
    a_1(T) &= \alpha_1 (T - T_{c,1}) \\
    a_2(T) &= \alpha_2 (T - T_{c,2})
\end{align}

\subsection{Parameters}
\begin{itemize}
    \item $T_{c,1} = 200$~K, $T_{c,2} = 250$~K
    \item $\alpha_1 = 4.0$, $\alpha_2 = 1.0$
    \item $b_1 = b_2 = 1.0$
    \item $\gamma = 1.0$ (positive: phases compete, if zero, no competition between phases, and they coexist.)
\end{itemize}

The $\gamma$ term has its physical origin in the lattice distortion incompatibility between both order parameters, and hence it is the driver of the phase competition and originates from the first-order transition.
Note that the interaction renormalizes the transition temperature for the $q_\mathrm{CO}$, lowering it and creating the first-order phase transition, as shown in the figure with the results.

\subsection{Results}
Minimization of $F$ with respect to $\psi_\mathrm{CO}$ and $\psi^{*}$ at each temperature shows:
\begin{itemize}
    \item For $T > 250$~K: Normal (3$\times$1) phase.
    \item For 180~K $<$ T $<$ 250 K: q$^*$ CDW dominates.
    \item For T $<$ 180~K: q$_\mathrm{CO}$ CDW becomes stable, suppressing q$^*$.
\end{itemize}

\begin{figure}[h!]
\centering
\includegraphics[width=0.8\textwidth]{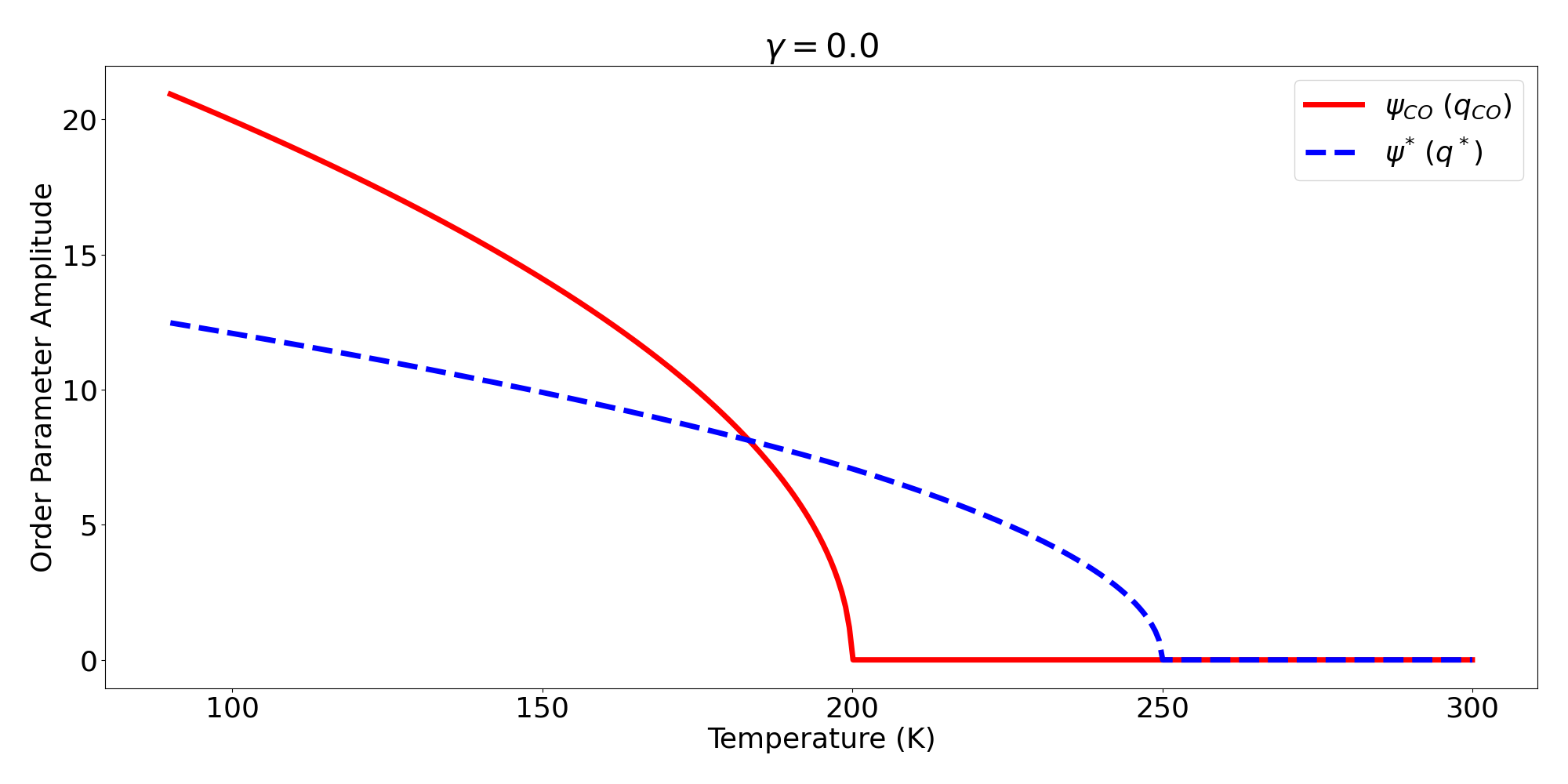}
\includegraphics[width=0.8\textwidth]{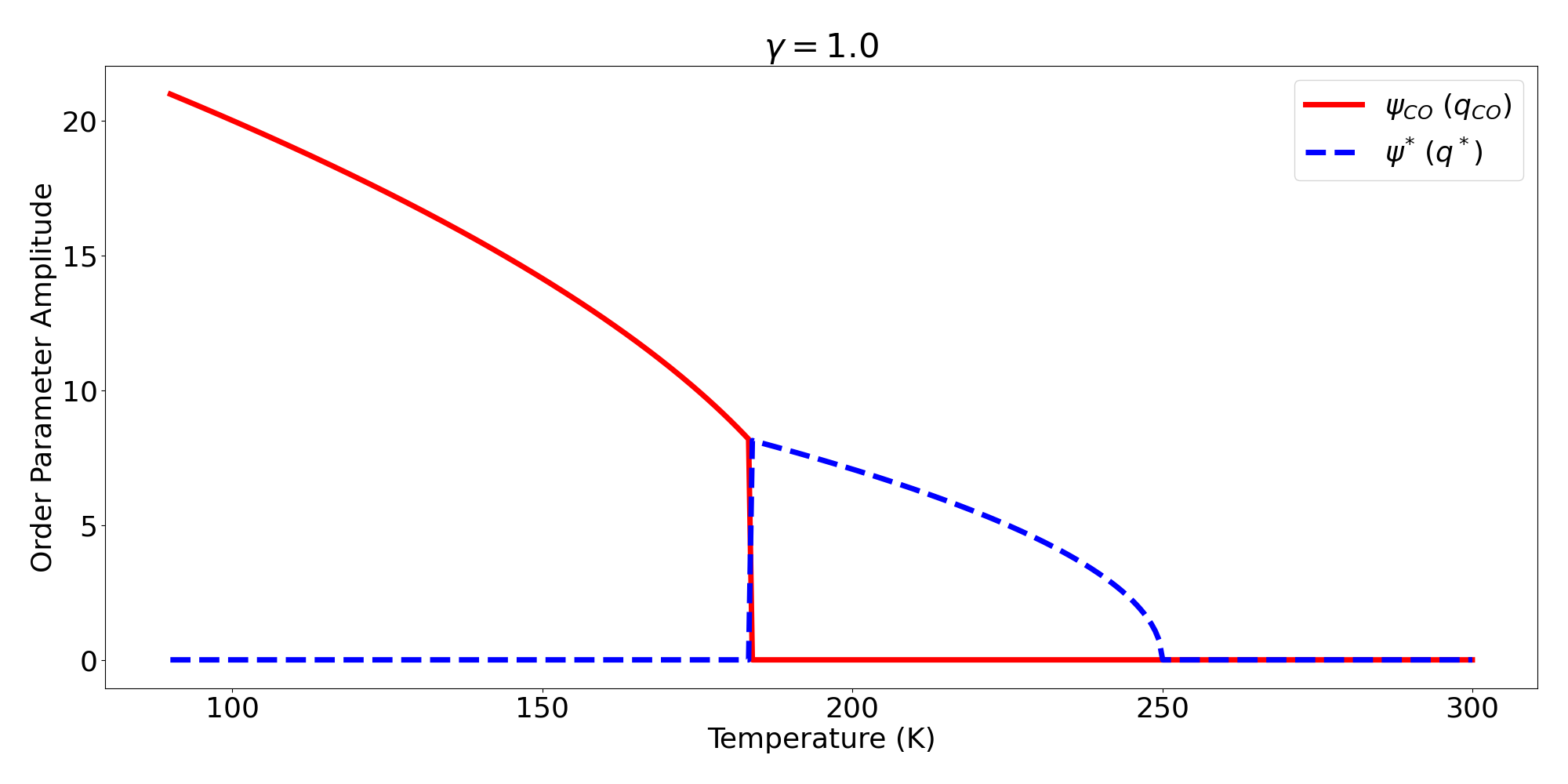}
\caption{Temperature dependence of the order parameters $\psi_\mathrm{CO}$ and $\psi^{*}$. The $q^*$ CDW condenses first, followed by a first-order transition to $q_\mathrm{CO}$.}
\end{figure}

\bibliography{TaTe2_References}